\documentclass{aa}  

\usepackage{graphicx}
\usepackage{subcaption}
\usepackage{txfonts}
\usepackage{ifthen}
\usepackage{natbib}
\usepackage{amssymb, amsmath}
\usepackage{cancel} 
\usepackage{booktabs} 
\usepackage{appendix}
\usepackage{etoolbox}
\usepackage[T1]{fontenc}
\usepackage{paralist}
\usepackage{commath} 
\usepackage[
bookmarks=true,           
bookmarksnumbered=true,   
colorlinks=true,          
citecolor=blue,           
linkcolor=blue,           
menucolor=blue,           
urlcolor=blue,            
linkbordercolor={0 0 1},  
pdfborder={0 0 1},
frenchlinks=true]{hyperref}
\newcommand\rsun{\ifmmode{R\sb{\odot}}\else$R\sb{\odot}$\fi}
\newcommand\mearth{\ifmmode{M\sb{\oplus}}\else$M\sb{\oplus}$\fi}
\newcommand\rearth{\ifmmode{R\sb{\oplus}}\else$R\sb{\oplus}$\fi}

\renewcommand{\pd}[2]{\frac{\partial {#1}}{\partial {#2}}}
\newcommand{\dfracpd}[2]{\dfrac{\partial {#1}}{\partial {#2}}} 

\newcommand{\vek}[1]{\boldsymbol{#1}} 
\newcommand{\vekvrel}{ {\vek{v}_{\mathrm{rel}}} } 
\newcommand{\mfp}{\lambda}
\newcommand{\mfpgas}{\mfp_{\mathrm{g}}} 
\newcommand{\rhogas}{\rho_\mathrm{g}} 
\newcommand{\kB}{\mathrm{k_B}} 
\newcommand{\Tbg}{T_\mathrm{bg}} 
\newcommand{\Lstar}{L_\star} 
\newcommand{\Lsun}{L_\odot} 
\newcommand{\mstar}{M_{\star}} 
\newcommand{\rchar}{{r_\mathrm{c}}} 
\newcommand{\vispow}{\gamma} 
\newcommand{\stokes}{\mathrm{St}} 
\newcommand{\tstopp}{\tau_\mathrm{s}} 
\newcommand{\tptes}{\tau_\mathrm{ptes}} 
\newcommand{\tdrift}{\tau_\mathrm{dr}} 
\newcommand{\optdepth}{\tau} 
\newcommand{\optdepthR}{\tau_\mathrm{R}} 
\newcommand{\reynolds}{\mathrm{Re}} 
\newcommand{\stmin}{\stokes_\mathrm{min}} 
\newcommand{\stmax}{\stokes_\mathrm{max}} 
\newcommand{\vfrag}{v_\mathrm{f}} 
\newcommand{\Mdisk}{M_{\mathrm{disk}}} 
\newcommand{\Gnewt}{G} 
\newcommand{\Mstar}{M_\star} 
\newcommand{\aturb}{\alpha_\mathrm{t}} 
\newcommand{\press}{P} 
\newcommand{\surfp}{\Sigma_\mathrm{p}} 
\newcommand{\surfpeb}{\Sigma_\mathrm{peb}} 
\newcommand{\surfgas}{\Sigma_\mathrm{g}} 
\newcommand{\dotsurfp}{\dot{\Sigma}_\mathrm{p}} 
\newcommand{\pebflux}{\dot{M}_\mathrm{peb}} 
\newcommand{\Mcr}{\dot{M}_\mathrm{cr}} 
\newcommand{\msun}{M_\odot} 
\newcommand{\peff}{\varepsilon} 
\newcommand{\trapdist}{d} 
\newcommand{\tlife}{\tau_\mathrm{l}} 
\newcommand{\tform}{\tau_\mathrm{f}} 
\newcommand{\Lx}{L_\mathrm{X}} 
\newcommand{\AU}{\mathrm{au}} 
\newcommand{\yr}{\mathrm{yr}} 
\newcommand{\dustrad}{a} 
\newcommand{\vdrift}{v_\mathrm{drift}} 
\newcommand{\dustsurf}{\Sigma_{\mathrm{d}}} 
\newcommand{\numdens}{n} 
\newcommand{\nsize}{n_{\dustrad}} 
\newcommand{\mptes}{m_\mathrm{p}} 
\newcommand{\mgas}{ {m_{\mathrm{g}}} } 
\newcommand{\mdust}{m} 
\newcommand{\msol}{\mdust} 
\newcommand{\hscale}{h_\mathrm{g}} 
\newcommand{\hscdust}{h_\mathrm{d}} 
\newcommand{\Okepl}{\Omega} 
\newcommand{\cs}{c_\mathrm{s}} 
\newcommand{\cdrag}{C_{\mathrm{D}}} 

\newcommand{\mdens}{\rho} 
\newcommand{\rhoint}{\mdens_\mathrm{s}} 
\newcommand{\dtog}{Z} 
\newcommand{\rhog}{{\mdens_{\mathrm{g}}} } 
\newcommand{\rhod}{ \mdens_{\mathrm{d}} } 
\newcommand{\sound}{c_{\mathrm{s}}} 
\newcommand{\torb}{t_\mathrm{orb}} 
\newcommand{\vth}{v_\mathrm{th}} 
\newcommand{\vdustk}[1]{v_\mathrm{#1}} 
\newcommand{\diffd}{D_\mathrm{d}} 
\newcommand{\diffg}{D_\mathrm{g}} 
\newcommand{\diffdeff}{D_\mathrm{d,eff}} 

\newcommand{\vrel}{ {v_{\mathrm{rel}}} } 
\newcommand{\crosssec}{\sigma}
\newcommand{\crosssecHtwo}{\crosssec_\mathrm{H2}}
\newcommand{\vfdrag}{\vek{F}_\mathrm{D}} 
\newcommand{\fdrag}{F_\mathrm{D}} 
\newcommand{\molvisc}{\nu_{\mathrm{mol}}} 
\newcommand{\lpgrad}{\dfracpd{\ln{\press}}{\ln{r}}}
\newcommand{\sigmaSB}{\sigma_\mathrm{SB}} 

\newcommand{\ie}{{i.\,e.}}

\newcommand{\eg}{{e.\,g.}}

\usepackage{xcolor}

\newcommand{\std}[1]{{\color{gray}{#1}}}
\newcommand{\stdtwo}[1]{\mathbf{#1}}

\usepackage{cancel}

\begin{document}

   \title{Constraining the parameter space for the Solar Nebula}

   \subtitle{The influence of disk properties on planetesimal formation}


   \author{Christian~T.~Lenz
          \inst{1}\fnmsep\thanks{Member of the International Max Planck Research School for Astronomy and Cosmic Physics at the Heidelberg University}
          \and
          Hubert~Klahr\inst{1}
          \and
          Tilman~Birnstiel\inst{2}
          \and
          Katherine Kretke\inst{3}
          \and
          Sebastian Stammler\inst{2}
          }

   \institute{Max Planck Institute for Astronomy,              K{\"o}nigstuhl 17, D-69117 Heidelberg,            Germany\\
              \email{lenz@mpia.de, klahr@mpia.de}
         \and
             University Observatory, Faculty of Physics, Ludwig-Maximilians-Universit{\"a}t M{\"u}nchen, Scheinerstr. 1, D-81679 Munich, Germany
        \and
            Southwest Research Institute, 1050 Walnut Ave, Suite 300, Boulder, CO, 80302, USA
             }

   \date{A{\&}A, accepted}

 
  \abstract
   {If we want to understand planetesimal formation, the only data set we have is our own Solar System. It is particularly interesting as it is so far the only planetary system we know of that developed life. Understanding the conditions under which the Solar Nebula evolved is crucial in order to understand the different processes in the disk and the subsequent dynamical interaction between (proto-)planets, once the gas disk is gone.}
   {Protoplanetary disks provide a plethora of different parameters to explore. The question is whether this parameter space can be constrained, allowing simulations to reproduce the Solar System.}
   {Models and observations of planet formation provide constraints on the initial planetesimal mass in certain regions of the Solar Nebula. By making use of pebble flux-regulated planetesimal formation, we perform a parameter study with nine different disk parameters like the initial disk mass, initial disk size, initial dust-to-gas ratio, turbulence level, and more.}
   {We find that the distribution of mass in planetesimals in the disk depends on the planetesimal formation timescale and the pebbles' drift timescale. Multiple disk parameters can influence pebble properties and thus planetesimal formation. However, it is still possible to draw some conclusions on potential parameter ranges.}
   {Pebble flux-regulated planetesimal formation seems to be very robust, allowing simulations with a wide range of parameters to meet the initial planetesimal constraints for the Solar Nebula. I.e., it does not require a lot of fine tuning.}


   \maketitle
%

\section{Introduction}\label{introduction}
In order to form planets, tiny micron sized dust grains have to grow to hundreds or thousands of kilometers. First, grains grow by collisions with other grains. But at some point they 
cannot continue to grow because either (1) the relative velocities become so high that a collision leads to fragmentation \citep{BlumMuench1993,BlumWurm2008,GundlachBlum2014}, or (2) they start to drift faster toward the star, than they 
could potentially grow \citep{KlahrBodenheimer2006,Birnstiel2012}. There is also the bouncing barrier \citep{zsom2010}, but charging effects might lead to growth to to sizes an order of magnitude above this barrier's limit \citep{steinpilz2019}. Laboratory experiments point in the direction of low fragmentation speeds for icy particles \citep{musiolik2019} of around $1\,\mathrm{m\,s^{-1}}$. This could cause particles to hit the fragmentation barrier first, which is why we are not considering the bouncing in this paper.

It is believed that planets are formed by so-called planetesimals, of a few to hundreds of kilometers in size. These planetesimals are the building blocks of planets. Once they have formed, accretion of pebbles \citep{OrmelKlahr2010} may become important too \citep[for a review see, e.g.,][]{ormel2017}.
But since grains stop growing at some point, continuous growth leaves a missing link between roughly mm--dm to planetesimal diameters ($\sim100\,\mathrm{km}$) via continuous growth.


If there is a pressure bump in the disk, \eg, caused by a vortex or zonal flow, particles can get trapped around the center of the bump \citep{Whipple1972,BargeSommeria1995}. 
After the accumulation of enough pebbles, the streaming instability \citep{YoudinGoodman2005} can potentially be the dominant turbulent process to trigger fragmentation in the laminar case ($\aturb=0$), or in the turbulent case ($\aturb>0$) gravoturbulent planetesimal formation can occure \citep{johansen2006,johansen2007nat}.

To summarize, we follow the idea that pebbles form planetesimals in a gravoturbulent process leading to a Gaussian-like size distribution of planetesimals that peaks around $100\,\mathrm{km}$ in diameter \citep{KlahrSchreiber2015,andydiss}. These planetesimals can then build planetary embryos, which can grow via further accretion of both planetesimals and pebbles to form (proto-)planets.

The initial distribution of planetesimals is one of the biggest unknowns in planet formation models. The radial distribution in the disk is important for embryo formation and subsequent accretion of planetesimals onto embryos. While we can observe protoplanetary disks and debris disks around other stars, we cannot observe planetesimal populations. There is only one system where we have relatively good knowledge of the present day small body population --- the Solar System. This is also the system which has been modeled the most, so the strongest constraints we have on the initial planetesimal population comes from our Solar System, even though (as we will discuss) there are a lot of uncertainties. The so-called minimum mass Solar Nebula (MMSN) \citep{weidenschilling1977mmsn,Hayashi1981} is the most commonly used assumption, but it is not based on a modern understanding of planet formation. 

\cite{lenz2019} presented a pebble flux-regulated model for the planetesimal formation rate. In this model, a spatial planetesimal distribution evolves with time and leads to a physically motivated planetesimal density disk profile. So it is sensible to ask, can this model fit the constraints of our Solar System? And how finely tuned does this model have to be?

This paper is structured as follows. In section 2 we discuss possible constraints for initial planetesimals in different regions of the Solar Nebula. Section 3 describes the model we are using. Section 4 presents the results and we conclude in section 5.

\section{Mass Constraints for the Initial Planetesimal Population}\label{sec:mass_constr}
In this section we will review literature studies to infer constraints on the initial planetesimal mass in
various regions of the Solar Nebula. A summary of this literature research is depicted in Fig.~\ref{fig:mass_constraints_overview}.
\subsection{Mercury Region: Interior to $0.7\,\AU$}
There is no observed stable population of asteroids 
within the orbit of Mercury \citep{steffl2013}, even though e.g. \cite{campins1996} found a dynamically stable region in the Solar System's inner region. On top of that, no models for terrestrial planet formation require any planetesimals inside of Mercury’s orbit ($\sim0.4\,\AU$). Additionally, terrestrial planet models fit observations better if the planetesimal disk is truncated around 0.7 AU \citep{hansen2009,walsh2011,levison2015,morbidelli2016}.
Even though there are some models suggesting how to clear out this region \citep[e.g.][]{ida2008,batygin2015,volk2015}, none of these are demonstrated in 
a comprehensive way. 
However, this implies that we can’t define an upper limit on the mass of initial planetesimals in this region.
Low mass short-period planets around other stars may indicate that in other planetary systems there were planetesimals in short period orbits that formed planets via in
situ planetesimal accretion \citep[e.g.][]{chiang2013,hansen2012} or pebble accretion \citep[e.g.][]{chatterjee2013}, but the presence of migration implies that there 
is no evidence that a population of initial planetesimals was present within Mercury's orbit in our own Solar System.

\textbf{Bottom line}: 0 to an unknown upper limit

\subsection{Earth/Venus Region $0.7-1\,\AU$}
As it was shown by \cite{hansen2009}, placing $\sim2\,\mearth$ of oligarchs within $0.7-1\,\AU$ can lead to good matches to the sizes and spacing of the terrestrial planets of the Solar System. The results from the parameter study of \cite{kokubo2006}, in terms of the number of Earth-like planets within 0.5 to 1.5$\,\AU$, seem to be very robust with respect to mass and radial profile of oligarchs. They have shown that one can still obtain reasonable results with a total initial mass of $2.77\,\mearth$ in that region. For $\sim23\,\mearth$, Super-Earths form. 

Pebble flux can allow a significant amount of mass to be transported into this region. If there are enough pebbles and appropriate disk structure, it is possible to produce reasonable Solar System analogs beginning with $<10^{-2}\,\mearth$ planetesimal masses \citep{levison2015}. 
If Jupiter migrated inwards and then out of the Asteroid Belt \citep[known as the grand tack;][]{walsh2011}, to leave about the needed mass in the current Asteroid Belt it would have implanted $\sim1\,\mearth$ 
of material into the Earth/Venus forming region. This suggests that either primordially or after early pebble accretion there was initially $1\,\mearth$ of planetesimals/embryos in this region \citep{walsh2011}.
Since the grand tack removes most objects in the Asteroid Belt in simulations, the inital planetesimal population needs to be massive enough early on. This probably implies that pebble accretion only grew the mass
of the Asteroid Belt by a factor of a few at most, and thus the terrestrial planet region probably only grew by a factor of a few as well. As a rough estimate we can assume the inital planetesimal mass to be around
$0.1\,\mearth$.

\textbf{Bottom Line}: 0.1 (if there are enough pebbles that can be accreted) to $2.77\,\mearth$ (if there aren't) 

\subsection{Asteroid Belt: $2-3\,\AU$}
The Asteroid Belt currently has a mass of about $5\cdot10^{-4}\,\mearth$ \citep[e.g.][]{kresak1977}, where roughly 50\% of the mass is in the 4 largest objects, with 1/3 in Ceres. Over the 
history of the Solar System, it has potentially been depleted by the following effects.
(1) Over the last 4 Gyr dynamical chaos in the current structure of the Solar System has removed about $\sim50\%$ of the mass of the Asteroid Belt \citep{minton2010}.
Vesta's crust indicates that the Asteroid Belt population was only modestly larger than it is today at the time the mean collision velocities were pumped up to 5 km/s 
(i.e. the current mean impact velocity in the main belt region; \cite{bottke1994}). If the Asteroid Belt had substantially more 30 km-sized planetesimals in it over the last 4 Gyr 
than it has today, Vesta would be expected to have more than 1 large basin \citep{bottke2005sizeDistriMainBelt,bottke2005linking,OBrienGreenberg2005}.
If planetesimals were “born big” \citep{morbidelli2009}, i.e. $\gtrsim80\,\mathrm{km}$, this suggests that collisional evolution should not be particularly important in removing material; at least not more than a factor of a few.
The late Jupiter-Saturn interaction in terms of reshuffling of the Giant planets likely depleted the Asteroid Belt by a factor of $\sim2$ to $\sim10$ 
\citep{minton2010}. If the Grand Tack happened \citep{walsh2011}, then only few times $10^{-3}$ to a few times $10^{-4}$ of the population would have survived \citep[ignoring newly implanted plantesimals from other regions; see e.g. Fig. 7 in][]{morbidelli2015}. 
One should also note that 
in this model the C-complex asteroids, which comprise about $75\%$ of the asteroids \citep{gradie1989} and include Ceres, are implanted from the outer Solar System. But these 
modifications are swamped by the uncertainty in the clearing rate on the migration efficiency.
If pebble accretion plays a crucial role for the growth of large asteroids ($>200\,\mathrm{km}$ in diameter), then this also would reduce the “initial mass” of planetesimals that is needed. This effect probably would be a factor of $\sim2$ 
\citep{johansen2015}.

\textbf{Bottom line}: $\sim2\cdot10^{-3}\,\mearth$ (4 times current mass) to $\sim5\mearth$ 

\begin{figure*}[tb]
  \centering
  \includegraphics[width=0.8\linewidth]{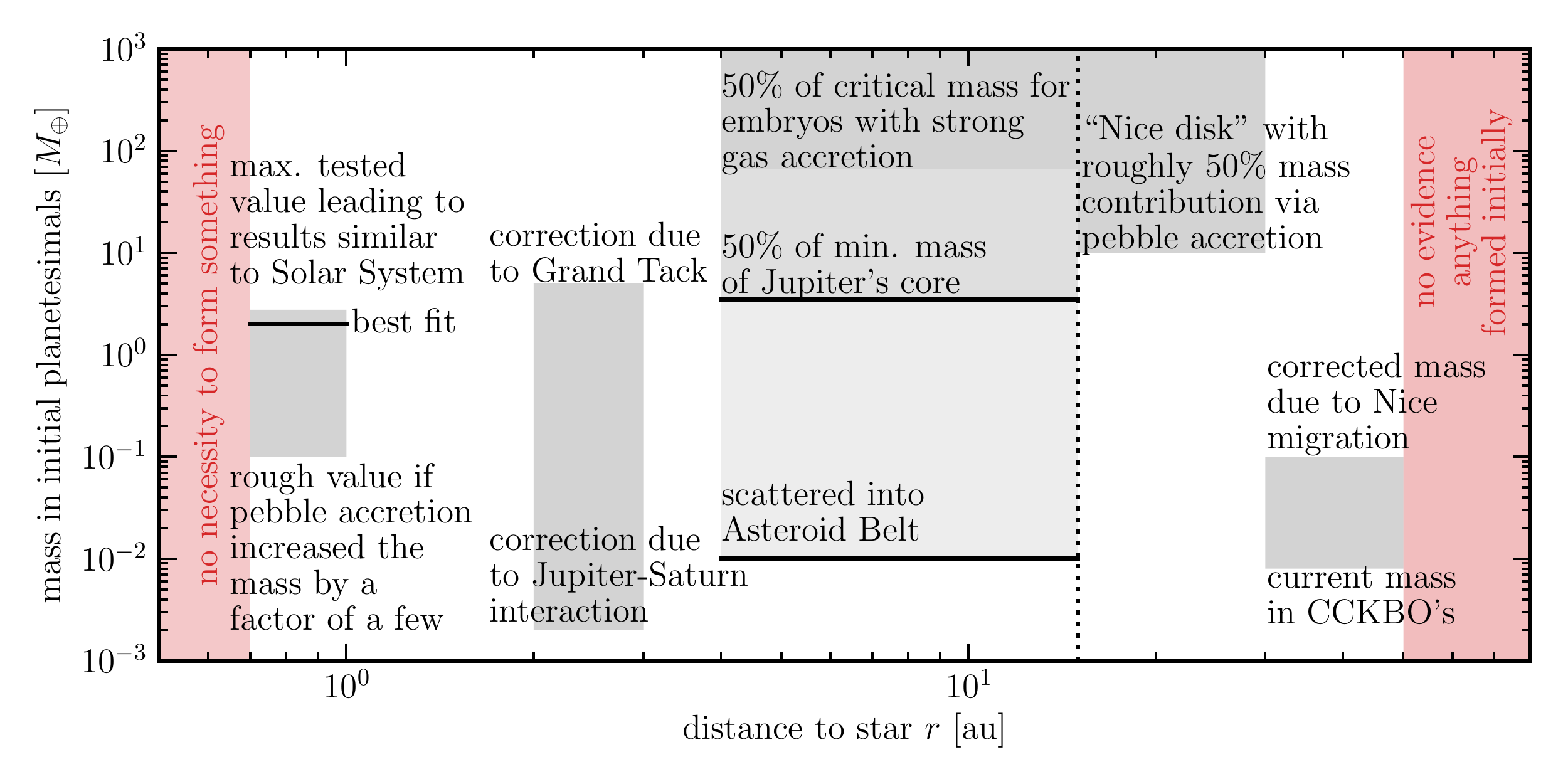}
  \caption
  {Graphical representation summarizing section~\ref{sec:mass_constr} on mass constraints of initial planetesimal masses in different regions of the Solar Nebula. Masses given by the values on the ordinate apply for the entire marked regions. Boxes that extend to the border of the plot indicate unknown upper limits. The dotted vertical line separates the two touching regions.}
  \label{fig:mass_constraints_overview}
\end{figure*}
\begin{figure}[thb]
  \centering
  \includegraphics[width=\linewidth]{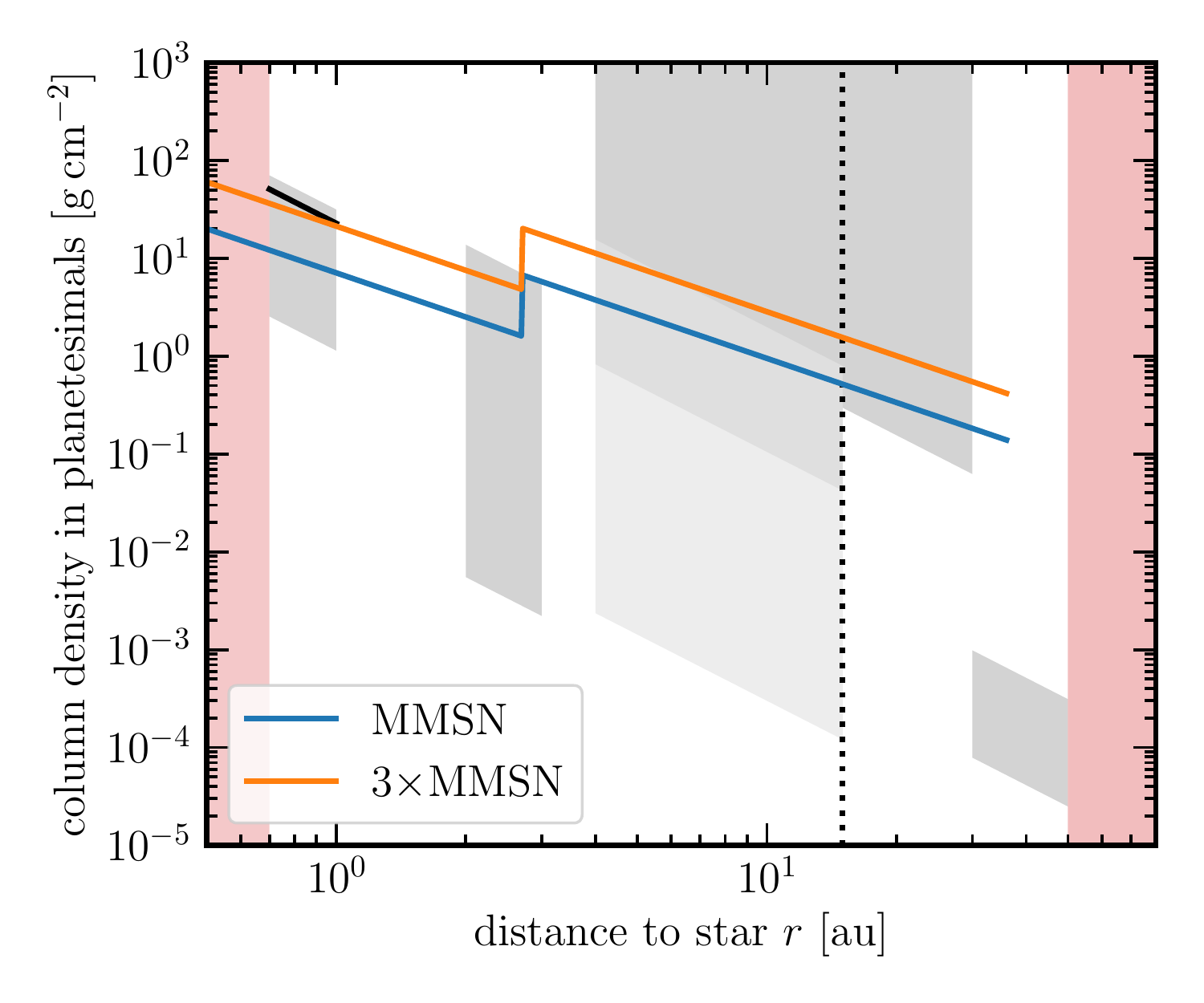}
  \caption
  {As Fig.~\ref{fig:mass_constraints_overview}, but converted into column densities, assuming a power-law shape $\propto r^{-2.25}$ (see Eq.~(37) of \cite{lenz2019} for a motivation). The blue (orange) line shows (three times) the minimum mass Solar Nebula profile for solids \citep{weidenschilling1977mmsn,Hayashi1981}.}
  \label{fig:density_constraints_overview}
\end{figure}
\subsection{Giant Planet Forming Region (possibly) $4-15\,\AU$}
This region could also have an inner (outer) edge that is further in (out). But this would not change the constraints dramatically.

The lower mass limit of Saturn's core is around $8-9\mearth$ \citep{saumon2004,helled2008}, Jupiter's core has at least $7\mearth$ \citep{wahl2017}. So far, we do not know how much of this mass was originally in planetesimals from the appropriate region. Hence, for the lower limit, we will ignore Saturn and take 50\% \citep[assuming that the other half stems from pebbles,][]{bordukat19thesis} of the lower mass estimate from \cite{wahl2017} which gives $3.5\mearth$. 

In order to reach critical masses for strong gas accretion, around 5 times the mass of the MMSN seems to be needed \citep[][these simulations used planetesimals of 20\,km in diameter]{ThommesDuncan2006}. 
This leads to $132\,\mearth$ within 4 to $15\,\AU$. Assuming that 50\% of the mass was contributed by pebbles \citep{bordukat19thesis}, the lower limit would be around $66\,\mearth$. This lower limit is still high, but desublimation effects just outside the ice line can lead to a pile-up in planetesimals by a factor of $\sim5$ \citep{drazkowska2017,schoonenberg2018}. This effect is not included in this work.

\cite{raymond2017} found that $\sim10\%$ of the asteroids around the giant planets were scattered into the asteroid belt. So in order to explain the mass of C-type asteroids in the Asteroid Belt, there probably had to have been around few times $10^{-2}\,\mearth$ of asteroids in the giant planet forming region. This value would be the absolute minimum for this region.

\textbf{Bottom line}: $\sim66\,\mearth$ to unknown high mass


\subsection{The Nice Disk $\sim15-30\,\AU$}
In order to match the observed structure of the Kuiper Belt, where many objects are in resonance with Neptune, outward migration of the giant planets is the preferred explanation. For this type of outward migration, planetesimal driven migration is the leading explanation \citep{fernandez1984,malhotra1995,thommes1999}. 
the Nice Model \citep{tsiganis2005,morbidelli2005,gomes2005} is a comprehensive model explaining how this outward planetesimal driven migration could have occurred.
In this model the giant planets initially formed closer to the sun than their current locations, and migrated outwards due to interactions with a planetesimal disk known as ``the Nice disk''. 
Even though the details of the model have been changed \citep[e.g.][]{morbidelli2007,levison2011}, a number of features in the small body reservoirs of our 
Solar System can be explained if this population did initially exist and the planets migrated through it. We give a few examples:
\begin{itemize}
 \item \emph{Jupiter's Trojans}: Calculations showed that a mass of around $\sim35\,\mearth$ can agree with the current population of the Jupiter’s Trojans \citep{morbidelli2005}. 
 With a newer variation, known as the Jumping Jupiter Model, good matches are found with planetesimal disk masses $\sim14-28\,\mearth$ \citep{nesvorny2013}.
 \item \emph{Kuiper Belt}: Models with grainy migration of Neptune with a disk mass of $\sim20\,\mearth$ (1000 Pluto sized objects) match the detailed characteristics of the objects in the 3:2 
 resonance \citep{nesvorny2016}.
 \item \emph{Comets}: Gas drag prevents km-sized planetesimals from being scattered into the Oort Cloud while the gas disk is still around \citep{brasser2007}. This suggests that 
 the long period comets were scattered into the Oort cloud after the gas disk went away, a natural outcome of something like the Nice Model. An initial population of around 
 $30\,\mearth$ is needed to populate the Oort Cloud \citep{dones2004}, however the existence of more massive planets in the inner Oort Cloud \citep[e.g. a planet 9,][]{batygin2016}
 could decrease the required reservoir size. However, since comets could have potentially been shared between stars in the birth cluster under favorable conditions, 
 it is possible comets are not a reliable constraint \citep{levison2010}.
 \item \emph{Ice giant ejection}: In the models in which an ice giant is ejected from the Solar System, the best overall structure of the Solar System, for example surviving terrestrial planets, needs 
 $\sim20\,\mearth$ in planetesimals \citep{nesvorny2012}.
\end{itemize}
Additionally, it has been found that the column density profile of planetesimals has a minimal effect on the outcomes for a relatively broad range of power-laws. This was tested by \cite{batygin2010} for $\surfp\propto r^{-k}$, where $k=1\dots 2$, with inner edge $r_\mathrm{in}\sim 12\,\AU$ and outer radius $r_\mathrm{out} = 30\,\AU$.
The Nice scattering occurred after the disk went away so that pebble accretion could have increased the total mass of the Nice disk. However, because the 
disk was very likely flaring in the outer regions, it is unlikely that pebble accretion was efficient and increased the total mass in these regions more than a factor of $\sim2$ 
\citep{lambrechts2012}.

\textbf{Bottom line}: $\sim10\,\mearth$ seems to be needed


\subsection{The Cold Classical Kuiper Belt $\sim30\,\AU-50\,\AU$}
There is a population of objects in the Kuiper belt with low eccentricities and inclinations which look as though they are not transplanted, though likely to be primordial. 
Observations indicate that the mass of the current classical population is $8\cdot10^{-3}\,\mearth$ \citep{fuentes2008}.

If the larger KBOs were formed by coagulation from small planetesimals, there must have been significantly more mass in this region in the early stages. 
For instance, \cite{pan2005} suggested that the high end size distribution could be matched by collisional evolution. 
However, if one combines more modern description laws with the need to preserve wide binaries, 
one cannot match the observed population in this type of collisional environment. 
Therefore, this suggests that planetesimals formed as large bodies and that the total mass of the cold classcial Kuiper Belt objects (CKBO), 
dominated by bodies larger than diameters of $\sim100\,\mathrm{km}$ has not evolved significantly \citep{nesvorny2011}.

The Nice migration may have dynamically depleted the Kuiper Belt by up to an order of magnitude \citep{morbidelli2008}. 
\cite{singer2019} found a lack of small craters on Pluto and Charon, indicating that planetesimals in the Kuiper Belt are not a collisionally evolved population, or that collisions destroyed small planetesimals. 

\textbf{Bottom line}: 0.008 to $\sim0.1\,\mearth$


\subsection{Beyond $50\,\AU$}
A radial distance of roughly $50\,\AU$ appears to be a real edge to the cold classical Kuiper belt \citep{jewitt1998,trujillo2001,fuentes2008}. 
If there is a population with similar 
size and albedo’s to the observed KBO at $60\,\AU$, it's mass cannot be more than $8\%$ of the observed KBOs, as otherwise it would have been detected \citep{fuentes2008}. 
There are small bodies with semi-major axes greater than $50\,\AU$ in the Solar System, but most of them are dynamically coupled to the giant planets, suggesting that they 
have been scattered into their large orbits. Hence, they do not represent primordial orbits.
A possible exception to the objects coupled to giant planets are the Sedna type objects 
\citep{brown2004}, but these objects are on highly eccentric orbits, suggesting that they were 
scattered to their current locations and only decoupled from the rest of the Solar System 
after being scattered outward, e.g. by the tidal influence of the Sun’s birth cluster 
\citep{brasser2006,brasser2007,kaib2008,brasser2012}.

\textbf{Bottom line}: No evidence that anything formed at these distances initially

\section{The Model}
We use a new python based version of the \cite{Birnstiel2010} code called \texttt{DustPy} \citep{StammlerBirnstiel}. This code allows us to compute the radial motion and growth of particles, as well as gas evolution. 
\texttt{DustPy} is a 1-d (radial) code with analytical vertical integration, solving the Smoluchowski equation \citep{Smoluchowski1916} for particle growth. For more details see \cite{Birnstiel2010}. 
In the following we are describing basics of the dust model, a simple accretion heating model. The sink term we have chosen for the gas due to photoevaporative winds is shown in Appendix~\ref{sec:photoevapModel}. 
\subsection{Basics}
For simplicity we assume spherical compact particles with mass $m=(4/3)\pi\rhoint\dustrad^3$, where $\rhoint$ is the material density and $\dustrad$ the particle radius. 
\cite{epstein1924} derived a friction force under the condition that $\dustrad\ll\mfpgas$ and $\vrel\ll\vth$ for spherical particles
\begin{align}
 \label{eq:Fepstein}
 \vfdrag=-\frac{4\pi}{3}\dustrad^2\rhogas\vth\vekvrel.
\end{align}
The drag force particles feel while moving through a fluid ($\dustrad\gg\mfpgas$) is
\begin{align}
 \label{eq:Fstokes}
 \vfdrag=-\frac{\cdrag}{2}\pi\dustrad^2\rhogas\vrel\vekvrel,
\end{align}
$\rhogas$ being the gas mass density and $\vrel$ the relative velocity to the gas. $\cdrag$ is called the \emph{drag coefficient}. 
This drag law was already expressed --- in the same form but with a constant drag coefficient --- by \cite{Newton1729} in section 2 and 7 of his second book, for the impact of air on the falling motion of hollow glass spheres, where inertia is dominant over viscous forces. 
The first formulation of this drag formula in the form $\fdrag=\dustrad^2\rhogas\vrel^2\cdot f(\reynolds)$ --- here $\cdrag$ is given by some function depending on the Reynolds number
\begin{align}
  \reynolds=2\dustrad\vrel/\molvisc,
\end{align}
with molecular viscosity $\molvisc$ --- was given by \cite{rayleigh1892}. 
The drag coefficient for $\reynolds\leq2\cdot10^5$ is given by \citep{Cheng2009}
\begin{align}
 \label{eq:dragcoeff}
 \begin{aligned}
   \cdrag=&\frac{24}{\reynolds}\left(1+0.27\cdot\reynolds\right)^{0.43}\\
		&+0.47\left[1-\exp{\left(-0.04\cdot\reynolds^{0.38}\right)}\right].
 \end{aligned}
\end{align}
The molecular viscosity for hard spheres, neither attracting nor repulsing, is roughly given by 
\begin{align}
 \label{eq:molvisc}
 \molvisc=\frac{1}{2}\vth\mfpgas
\end{align}
\citep[see his Eq.~(249)]{chapman1916}, where the gas mean free path is
\begin{align}
 \mfpgas=\frac{1}{\sqrt{2}}\frac{1}{\numdens_\mathrm{g}\crosssec_\mathrm{g}}
\end{align}
and we further assume that the geometrical cross section $\crosssec_\mathrm{g}$ is that of molecular hydrogen $$\crosssecHtwo=2\cdot10^{-15}\,\mathrm{cm}^2.$$ 
\cite{massey1933} pointed out that this classical approximation is good enough for helium and hydrogen over a large range of temperatures (see their 
table~III on p. 450), \ie, quantum mechanics is not required. For cold temperatures ($\sim10\,\mathrm{K}$) quantum mechanics is important, but these temperatures are 
typical for the outer disk, where the gas density is so low that particles are in the Epstein drag regime anyway.

We define the \emph{stopping time} as
\begin{align}
    \tstopp=m\vrel/\fdrag,
\end{align}
following \cite{Whipple1972}. 
According to Newton's second law it is 
\begin{align}
    \dot{v}_\mathrm{rel}=-\vrel/\tstopp,
\end{align}
\ie, $\tstopp$ is the time it takes for the velocity of the particle relative to the gas to be reduced 
from $\vrel$ to $\vrel/e$. 
How well particles are coupled to the gas is described by their Stokes number, which we define by
\begin{align}
\label{eq:stokes}
 \stokes:=\Okepl\tstopp,
\end{align}
where
\begin{align}
  \Okepl=\sqrt{\frac{\Gnewt\Mstar}{r^3}}
\end{align}
is the Keplerian frequency. 
Since the Stokes number is the ratio of the stopping time, over which particles couple to the gas, and the dynamical gas timescale, small values ($\stokes\ll1$) 
mean that particles are coupled to the gas. Large values ($\stokes\gg1$) indicate that particles are decoupled from the gas. 
I.e., particles are coupled to the gas motion in less than an orbit for $\stokes\ll1$, and large Stokes numbers ($\stokes\gg1$) would need many orbits to synchronize to the gas motion.
If the mean free path of gas molecules $\mfpgas$ is large enough, particles are in the Epstein drag regime. If $\mfpgas$ is small compared to the particle radius $\dustrad$, they 
are in the fluid regime. The transition between the two regimes occurs around\footnote{Following \cite{weidenschilling1977}, this condition can be obtained by setting either the 
stopping time in the Stokes drag law and the Epstein drag regime or the two drag forces equal and making use of Eq.~\eqref{eq:molvisc}.} $\mfpgas=4\dustrad/9$. The Stokes number is thus
\begin{align}
 \label{eq:stokes-all-regimes}
 \frac{\stokes}{\Okepl}=
 \begin{cases}
 \displaystyle\rhoint\dustrad/(\rhogas\vth) &,\;\text{Epstein}\ (\mfpgas\geq4\dustrad/9)\\
 \displaystyle8/(3\cdrag)\frac{\rhoint}{\rhogas}\frac{\dustrad}{\vrel} &,\;\text{Fluid},\ \reynolds\leq2\cdot10^5\\
  \end{cases}
 .
\end{align}
If the fluid regime is reached, we follow \cite{Birnstiel2010} and assume the Stokes drag law regime, \ie, 
\begin{align}
    \cdrag=24/\reynolds 
\end{align}
for $\reynolds\leq1$ \citep[his Eq.~(126)]{Stokes1851}. 
If $\mfpgas<4\dustrad/9$, this leads to 
\begin{align}
    \stokes=\frac{2}{9}\frac{\rhoint}{\rhogas}\frac{\dustrad^2\Okepl}{\molvisc}.
\end{align}
This way the velocity of particles, their relative 
velocity to the gas, and the Stokes number don't have to be solved iteratively together. 
\subsection{Column Densities}
We define the column density as the mass per 3-dim volume, density $\rho$, vertically integrated over height $z$ of the disk,
\begin{align}
  \Sigma_i:=\int_{-\infty}^{\infty}\rho_i\,\dif{z}=2\int_{0}^{\infty}\rho_i\,\dif{z}
\end{align}
where $i=\{\mathrm{d,g,p}\}$ can be dust ($\mathrm{d}$), gas ($\mathrm{g}$), and planetesimals ($\mathrm{p}$). We use $\dustsurf$ as the column density, 
including all solid particles without planetesimals. If it has $\stokes$ as an argument, it is the column density of particles with this Stokes number. 
Following \cite{Birnstiel2010}, we define the dust column density distribution per logarithmic bin of grain radius $\dustrad$ as
\begin{align}
  \sigma_\mathrm{d}(r,\dustrad):=\int_{-\infty}^{\infty}\nsize(\dustrad,r,z)\msol(\dustrad)\dustrad\,\dif{z},
\end{align}
where $\nsize$ is the number density per grain size bin. This way, knowledge of the used size grid is not needed and the total dust column density is given by
\begin{align}
  \dustsurf(r)=\int_{-\infty}^{\infty}\sigma_\mathrm{d}(r,\dustrad)\,\dif{\ln{\dustrad}}.
\end{align}

As initial condition for the gas we use the self-similar profile \citep{lbp1974}
\begin{equation}
 \label{eq:gasini}
 \surfgas(r)=\frac{(2-\vispow)\Mdisk}{2\pi\rchar^2(1+\dtog_0)}\left(\frac{r}{\rchar}\right)^{-\vispow}\exp{\left[-\left(\frac{r}{\rchar}\right)^{2-\vispow}\right]},
\end{equation}
where $\dtog_0=\dustsurf/\surfgas$ is the initial solid-to-gas ratio in terms of column densities. The initial dust column density is then given by $\dtog_0\surfgas(t_0)$.

\subsection{Drift velocities}
Particles, from tiny dust grains up to boulders, are embedded in the gas disk. 
With the force of gravity from the central star balanced by the centrifugal force, particles move on Keplerian Orbits. The action of the gas pressure gradient on the particles can be neglected because the internal density of the particles is so much larger than the gas density. 
The gas does feel gravity, centrifugal force, and the pressure gradient force. If these forces balance each other, the gas moves on 
slightly sub-Keplerian orbits. Particles with $\stokes\lesssim1$ are coupled to the gas, thus they feel a centrifugal deficiency due to sub-Keplerian gas motion and drift radially inward. As long as $\stokes<1$ this 
leads to a stronger radial drift for increasing $\stokes$. If the particle Stokes number is larger than unity, they decouple from the gas and feel a headwind from the surrounding gas. 
The mass-to-surface ratio increases with size and this effect becomes weaker for increasing $\stokes$ ({\ie} increasing stopping time $\tstopp$). The steady-state solution for radial drift reads \citep{Nakagawa1986} 
\begin{align}
 \label{eq:radialdrift}
 \vdrift=\frac{\stokes}{\stokes^2+(1+\rhod/\rhog)^2}\frac{\hscale}{r}\lpgrad\sound
\end{align}
which reduces to 
\begin{align}
 \label{eq:radialdriftSingle}
 \vdrift=\frac{\stokes}{\stokes^2+1}\frac{\hscale}{r}\lpgrad\sound
\end{align}
for low dust-to-gas ratios \citep{weidenschilling1977}. We use the latter expression for this paper to save computation time.
\subsection{Planetesimal Formation Rate}
\label{sec:ptesFormRate}
For the planetesimal formation rate we follow \cite{lenz2019}. 
The model is based on the idea that pebble traps appear and disappear on a given timescale. 
In those pebble traps, pebble clouds can then collapse to planetesimals. 

In this model, the pebble flux (in mass per time)
\begin{align}
    \label{eq:pebflux}
	\pebflux:=2\pi r\sum_{\stmin\leq\stokes\leq\stmax}\abs{\vdrift(r,\stokes)}\dustsurf(r,\stokes).
\end{align}
is transformed into planetesimals over a \emph{conversion length} $\ell$:
\begin{align}
\label{eq:dotsurfp}
 \dotsurfp(r)=\frac{\pebflux}{2\pi r\ell}=\peff\frac{\pebflux}{2\pi r\trapdist}. 
\end{align}
We assume that this conversion length is proportional to the gas pressure scale height $\hscale$. Mass conversion from pebbles to planetesimals according to this recipe is only allowed if the condition
\begin{align}
	\label{eq:cond_ptes}
	\peff\tlife\pebflux>\mptes.
\end{align}
is fulfilled, where $\mptes$ is the mass of a single $100\,\mathrm{km}$ planetesimal and $\tlife$ is the lifetime of traps. In this paper, we assume that $\tlife=100\,\torb$ for all simulations. $\peff$ is the efficiency with which pebbles are transformed into planetesimals. For more details we refer to \cite{lenz2019}. With help of the mean radial trap separation $\trapdist$, one can relate this parameter to the conversion length, $\ell=\trapdist/\peff$. 

The jump from pebble-size to objects $100\,\mathrm{km}$ in diameter is a direct result of 
the particle diffusion timescale within the particle cloud and the collapse timescale \citep{KlahrSchreiber2015,andydiss,gerbig2020}. 

\subsection{Comparison to other Planetesimal Formation Rate Models}
\begin{figure}[t]
  \centering
  \includegraphics[width=\linewidth]{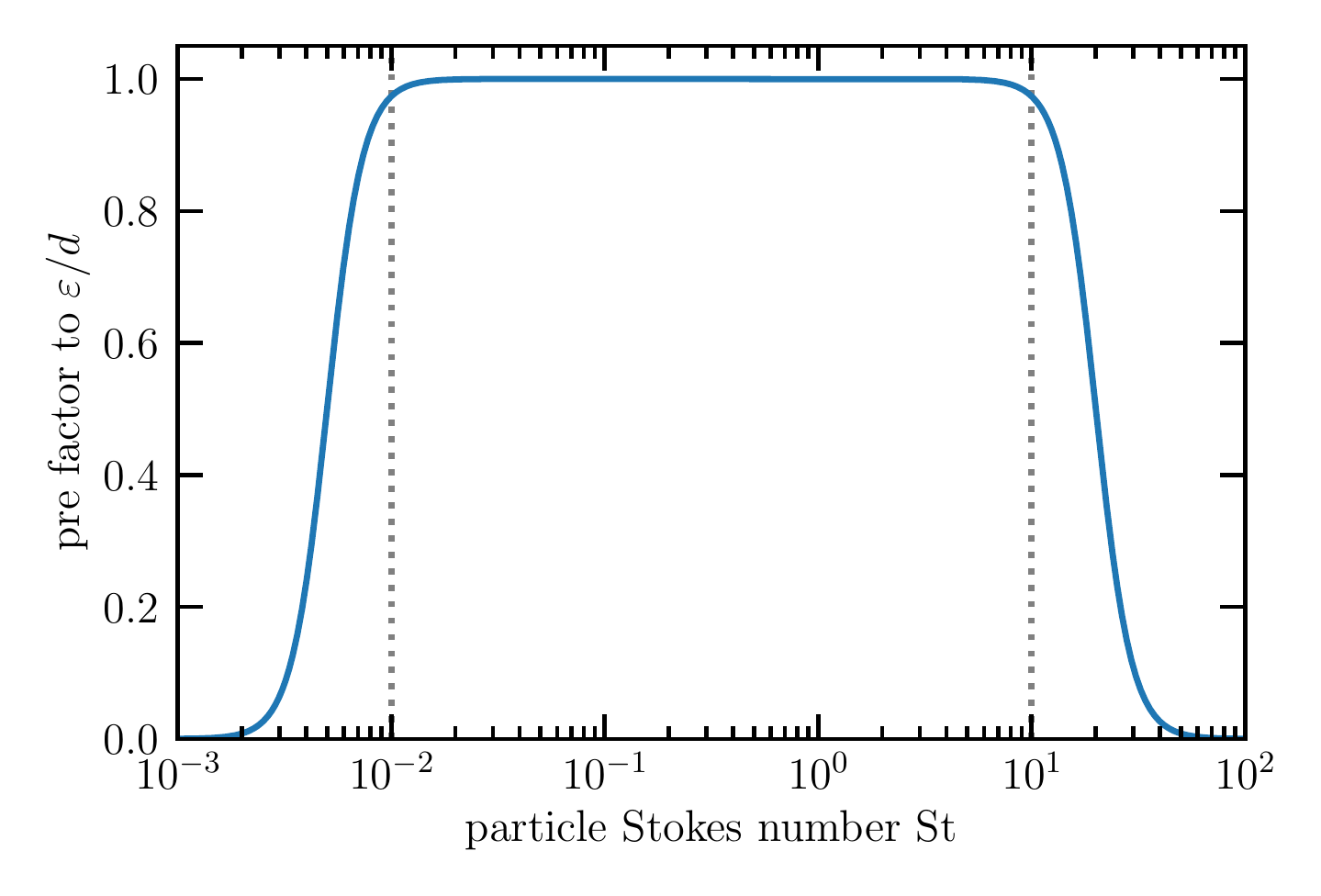}
  \caption
  {Illustration of Eq.~\eqref{eq:prefacEps} for $\stmin=0.01$ and $\stmax=10$.}
  \label{fig:prefacEps}
\end{figure}
The model for the planetesimal formation rate from \cite{lenz2019} differs from other models. In \cite{lenz2019} planetesimal formation is regulated by a conversion length scale, over which drifting particles are converted into planetesimals. The conversion length scale depends on the radial density of pebble traps and the efficiency of concentrating particles and converting pebble clouds into bound objects. \cite{drazkowska2016} and \cite{schoonenberg2018} suggest models for which planetesimal formation occurs with a certain efficiency per orbit from the local particle density. These models do assume that particles are not trapped while drifting. Potentially, an equivalent situation could be reached for explicit traps that build up and vanish on a given timescale everywhere in the disk with some average radial distance between each other. \cite{lenz2019} parameterized this via the conversion length $\ell$, see Eq.~(3) in their paper. 

Adding a gas gap to the simulation, \cite{stammler2019} used the model of \cite{schoonenberg2018} to produce planetesimals just outside this gap and were able to reproduce the observed optical depth of HD 163296.

\cite{eriksson2020} used the criterion from \cite{Yang2017} and assumed that all the available local mass is transformed into planetesimals once the condition is met for which particles in the midplane can concentrate to a particle-to-gas mass ratios of more than 10.

For further discussion of other planetesimal formation models see e.g. section 5.2 of \cite{lenz2019}.

\subsection{Advection-Diffusion Equation}
\begin{table*}[t]
\caption{Parameters that are checked in this study. Standard values are marked in gray. Those for the second fiducial set are shown in bold.  
Disk mass $\Mdisk$, characteristic radius $\rchar$, 
viscosity power-law index $\vispow$, which is also the power-law index of the column density of our 
initial condition for $r\ll\rchar$, breakup speed of grains $\vfrag$, initial dust-to-gas ratio $\dtog_0$, 
trap formation time $\tform$, turbulent viscosity parameter $\aturb$, planetesimal formation efficiency $\peff$, and X-ray luminosity $\Lx$. For comparison, the disk mass of the MMSN is roughly $0.013\msun$. In a separate row, we highlight our most appealing case, which includes a simple model for accretion heating that is not used in any other simulation.
} 
\label{tab:paramOverview}
\centering
\begin{tabular}{cccccccccc}
 \toprule\toprule
 \\
 $\Mdisk$ [$\msun$]  	& $\rchar$ [$\AU$]    & $\vispow$  & $\vfrag$ [$\mathrm{cm\,s^{-1}}$]   &  $\dtog_0$   & $\tform$ [$\torb$]  & $\aturb$ & $\peff$ & $\Lx$ [$\mathrm{erg\,s^{-1}}$]
 \\
 \midrule
 $0.02$         & $10$      & 0.5       & $\stdtwo{100}$        & 0.001   & 0     & $10^{-5}$ & 0.01 & $0$ 
 \\
 \std{$0.05$}     & \std{$\stdtwo{20}$}      & 0.8       & 300        & 0.003     & \std{300}   & $\stdtwo{10^{-4}}$ & $\stdtwo{0.03}$ & $3\cdot10^{28}$
 \\
 $\stdtwo{0.10}$         & $35$      & \std{$\stdtwo{1}$}         & \std{1000}       & \std{$\stdtwo{0.0134}$}     & 600   & \std{$10^{-3}$} & \std{0.1}  & $10^{29}$
 \\
                & $50$      & 1.2       &            &  0.02    & $\stdtwo{1000}$  & $10^{-2}$ & 0.3  & $3\cdot10^{29}$
 \\
                & $100$     & 1.5       &            &  0.03    &       &           & 1    & \std{$\stdtwo{10^{30}}$}
 \\
                &           &           &            &          &       &           &      & $10^{31}$
 \\
 \midrule
 0.10           & 20        & 1         &  200       & 0.0134   & 1600  & $3\cdot10^{-4}$ & 0.05 & $3\cdot10^{29}$
 \\
 \bottomrule
\end{tabular}
\end{table*}

\begin{table*}[t]
\caption{Parameters that can be excluded to reproduce the Solar Nebula based on Figures~\ref{fig:sim_sample1}, \ref{fig:sim_sample2}, \ref{fig:sim_sample_mass}, and \ref{fig:sim_sample_mass2}. We cannot exclude values for $\Lx$ within the range we have checked. For all the simulations presented in this table we used $\Lx=10^{30}\,\mathrm{erg\,s^{-1}}$. Here we concentrate on three constraints only: (1) the Cold Classical Kuiper Belt (CCKB) mass constraints, (2) the necessary mass to fulfill the Nice disk condition, (3) and the minimum mass solar nebula (MMSN) mass. In each row we highlight in bold those parameters that deviate from the default values.
} 
\label{tab:paramExclude}
\centering
\begin{tabular}{cccccccccc}
 \toprule\toprule
 \\
 $\Mdisk$ [$\msun$]  	& $\rchar$ [$\AU$]    & $\vispow$  & $\vfrag$ [$\mathrm{cm\,s^{-1}}$]   &  $\dtog_0$   & $\tform$ [$\torb$]  & $\aturb$ & $\peff$  & reason
 \\
 \midrule
 $0.05$         & $\mathbf{\geq35}$      & $1$       & 1000        & 0.0134   & 300     & $10^{-3}$ & 0.1 & fails CCKB condition
 \\
 $0.05$         & $20$      & $\mathbf{1.5}$       & 1000        & 0.0134   & 300     & $10^{-3}$ & 0.1 & fails CCKB condition
 \\
 $0.05$         & $20$      & 1       & 1000        & $\mathbf{\leq0.03}$   & 300     & $10^{-3}$ & 0.1 & <MMSN; fails Nice disk condition
 \\
 $0.05$         & $20$      & 1       & 1000        & 0.0134   & $\mathbf{<300}$     & $10^{-3}$ & 0.1 & fails CCKB condition
 \\
 $0.05$         & $20$      & 1       & 1000        & 0.0134   & 300     & $\mathbf{10^{-2}}$ & 0.1 & fails CCKB condition
 \\
 $0.05$         & $20$      & 1       & 1000        & 0.0134   & 300     & $10^{-3}$ & $\mathbf{\leq0.03}$ & <MMSN (for $\peff\lesssim0.01$); 
 \\
                &           &         &             &          &         &           &                     & fails Nice disk cond.
  \\
 $0.05$         & $20$      & 1       & 1000        & 0.0134   & 300     & $10^{-3}$ & $\mathbf{\gtrsim0.3}$ & fails CCKB condition
 \\
 \midrule
 $\mathbf{0.02}$         & $20$      & 1       & 100        & 0.0134   & 1000     & $10^{-4}$ & 0.03 & <MMSN; fails Nice disk condition
 \\
 $\mathbf{0.05}$         & $20$      & 1       & 100        & 0.0134   & 1000     & $10^{-4}$ & 0.03 & <MMSN; fails Nice disk condition
 \\
 $0.1$         & $\mathbf{\leq10}$      & 1       & 100        & 0.0134   & 1000     & $10^{-4}$ & 0.03 & fails Nice disk condition
 \\
 $0.1$         & $\mathbf{\geq35}$      & 1       & 100        & 0.0134   & 1000     & $10^{-4}$ & 0.03 & fails CCKB condition
 \\
 $0.1$         & $20$      & $\mathbf{0.5}$       & 100        & 0.0134   & 1000     & $10^{-4}$ & 0.03 & <MMSN
 \\
 $0.1$         & $20$      & $\mathbf{\geq1.2}$       & 100        & 0.0134   & 1000     & $10^{-4}$ & 0.03 & fails CCKB condition
 \\
 $0.1$         & $20$      & 1       & $\mathbf{<100}$        & 0.0134   & 1000     & $10^{-4}$ & 0.03 & <MMSN
 \\
 $0.1$         & $20$      & 1       & 100        & $\mathbf{\leq0.003}$   & 1000     & $10^{-4}$ & 0.03 & <MMSN; fails Nice disk condition
 \\
 $0.1$         & $20$      & 1       & 100        & 0.0134   & $\mathbf{0}$     & $10^{-4}$ & 0.03 & fails CCKB condition
 \\
 $0.1$         & $20$      & 1       & 100        & 0.0134   & 1000     & $\mathbf{\geq10^{-3}}$ & 0.03 & <MMSN; fails almost every cond.
 \\
 $0.1$         & $20$      & 1       & 100        & 0.0134   & 1000     & $10^{-4}$ & $\mathbf{\leq0.01}$ & <MMSN; fails Nice disk condition
 \\
 $0.1$         & $20$      & 1       & 100        & 0.0134   & 1000     & $10^{-4}$ & $\mathbf{\gtrsim0.3}$ & fails CCKB condition
 \\
 \bottomrule
\end{tabular}
\end{table*}

\begin{table*}[tbh]
\caption{Parameter ranges that could work for reproducing the Solar System. 
} 
\label{tab:paramRanges}
\centering
\begin{tabular}{lll}
 \toprule\toprule
 \\
 Symbol & Meaning & Comments
 \\
 \midrule
 $\Mdisk$                 & total disk mass  
 & $\Mdisk\gtrsim0.1\msun$ for $\vfrag\sim1\,\mathrm{m\,s^{-1}}$ and $\Mdisk\gtrsim0.02\msun$ for $\vfrag\gtrsim10\,\mathrm{m\,s^{-1}}$
 \\
 $\rchar$                   & char. radius
 &  $\lesssim50\,\AU$
 \\
 $\vispow$                          & initial inner column dust
 & For $\vfrag\gtrsim10\,\mathrm{m\,s^{-1}}$ $\vispow\sim0.5-1$. For $\vfrag\gtrsim1\,\mathrm{m\,s^{-1}}$ and $\Mdisk\gtrsim0.1\msun$, 
 \\
 & and gas density power-law index & $\vispow\sim0.5$ could work but $\vispow\sim1$ seems more likely
 \\
 $\vfrag$   & frag. speed
 & $\gtrsim1\,\mathrm{m\,s^{-1}}$ to allow pebbles with $\stokes\gtrsim10^{-2}$ to form
 \\
 $\dtog_0$                          & initial dust-to-gas ratio
 & $0.01\lesssim Z_0\lesssim0.03$ works more or less equally well, whereas $Z_0\lesssim0.003$ fails
 \\
 $\tform$                 & trap formation time
 & Traps needed at least $300\,\torb$ to form outside of $50\,\AU$ or never formed there
 \\
 $\aturb$                           & turbulence parameter
 & $\aturb\sim10^{-5}-10^{-3}$ (or only up to a few $10^{-4}$ if $\vfrag\sim1\,\mathrm{m\,s^{-1}}$)
 \\
 $\peff$                            & planetesimal formation efficiency
 & $0.002<\peff\hscale/\trapdist\lesssim0.06$ (if $\peff$ and $\trapdist/\hscale$ are constant)
 \\
 $\Lx$     & X-ray luminosity
 & For $\rchar\lesssim20\,\AU$, photoevaporation does not affect the final planetesimal profile 
 \\
 & & 
 significantly
 \\
 \bottomrule
\end{tabular}
\end{table*}

The particle diffusion coefficient $\diffd$ for species $i$ can be estimated with help of the gas diffusion coefficient
\begin{align}
    \label{eq:DgasDiff}
    \diffg=\aturb\sound\hscale
\end{align}
as \citep{YoudinLithwick2007}
\begin{align}
    \diffd^i=\frac{\diffg}{1+\stokes_i^2}.
\end{align}
This means small particles are diffused with the gas and larger particles are less influenced by gas diffusion. 
As first described by \citet[p. 66]{fick1855}, reviewed in more modern notation by \citet[his Eq. (1)]{Tyrrell1964}, and derived from fundamental principles by \citet[his Eq. (25)]{reeks1983}\footnote{Note that in some works the diffusive flux is written in a form where it is proportional to the gradient of "concentration", which is meant in the sense of mass per volume---not mass over mass fraction. In our notation this is given by the density $\mdens$.} 
the diffusive flux is given by
\begin{align}
    \label{eq:fick}
    \vek{J}_\mathrm{diff}^i = -\diffd^i\nabla\rhod^i
\end{align}
(see also \cite{cuzzi1993}) which gives the $z$-integrated version in radial direction
\begin{align}
    \label{eq:diff_flux_r}
    j_\mathrm{diff,r}^i = -\int_{-\infty}^{\infty}\diffd^i\pd{\rhod^i}{r}\dif{z}
    \approx -\diffd^i\pd{\dustsurf^i}{r}.
\end{align}
In the last step we used the fact that, due to the Gaussian shape of $\rhod^i$ in $z$-direction, the highest contribution of the integral comes from the region within $[-\hscdust^i,\hscdust^i]$, within which the gas density does not change by much (especially because $\hscdust^i<\hscale$). 
If the gas density is roughly constant, the particles Stokes number also stays roughly constant. If at $z=\hscdust^i$ the temperature is similar to the one of the midplane, $\diffd^i$ can be considered to be $z$-independent. As long as these conditions are met, the right hand side of Eq.~\eqref{eq:diff_flux_r} gives a good approximation. 
Since particles exhibit diffusive mixing due to turbulent gas motion, they are not able to move faster than the turbulent gas motion driving it. This maximum diffusion speed can be estimated to be \citep{Cuzzi2001}
\begin{align}
    v_\mathrm{max}\simeq\sqrt{\aturb}\sound.
\end{align}

We would like to point out that in the expressions in {\eg} \cite{Desch2017}, which are based on \cite{MorfillVoelk1984},
the diffusive flux is \emph{not} given by 
\begin{align}
    \nonumber
    j_\mathrm{diff,r}^i \neq -\diffd^i\surfgas\pd{}{r}\left(\frac{\dustsurf^i}{\surfgas}\right)
\end{align}
This expression is strictly speaking only valid for small particles, --- which couple to the gas motion on timescales shorter than the correlation time of the fastest turbulent eddy $\tau_\mathrm{Kolmogorov}$ , i.e.\ the smallest eddy at the dissipation scale of turbulence (Kolmogorov scale) $\tstopp<\tau_\mathrm{Kolmogorov}$ --- or for constant gas densities. Otherwise, Eq.~\eqref{eq:diff_flux_r} should be used. Unfortunately, we do not know the value of $\tau_\mathrm{Kolmogorov}$. 
For further details on the different turbulence regimes see e.g. \cite{OrmelCuzzi2007}. 
The difference between the two diffusion terms can be significant if the gas density drops quickly, as is the case for gap opening due to photoevaporation. 
Despite \cite{Dubrulle1995} using this diffusive flux for the vertical direction ($z$ derivative instead of $r$ derivative), their result for the particle scale height is still valid since the gas density does not change much in the vertical direction within one particle scale height.
By making use of Eq.~\eqref{eq:diff_flux_r}, the advection-diffusion equation reads
\begin{align}
\begin{aligned}
	&\pd{\dustsurf^i}{t}+\frac{1}{r}\pd{}{r}
    \left\{ 
    	r\left[  
        		\dustsurf^i\vdustk{r}^i-\diffdeff^i\pd{\dustsurf^i}{r}
         \right]
    \right\}
    \\
    &=-\frac{f_{\stokes}\peff}{\trapdist}\abs{\vdrift^i}\dustsurf^i
       \cdot\theta(\pebflux-\Mcr)
    .
\end{aligned}
\end{align}
Here, $\theta(\cdot)$ is the Heaviside function and 
\begin{align}
    \Mcr:=\frac{\mptes}{\peff\tlife}
\end{align}
is the critical pebble flux to allow planetesimal formation \citep{lenz2019}. 
We introduce a smoothing function for the Stokes number dependency of the efficiency parameter $\peff$
\begin{align}
    \begin{aligned}
    \label{eq:prefacEps}
    f_{\stokes}=\left\{\left[\exp{\left(-12\cdot\left(\lg{(\stokes)}-\lg{(\stmin/2)}\right)\right)}+1\right]\right.
    \\
    \times\left.\left[\exp{\left(12\cdot\left(\lg{(\stokes)}-\lg{(2\stmax)}\right)\right)}+1\right]\right\}^{-1}.
    \end{aligned}
\end{align}
This pre factor is displayed in Fig.~\ref{fig:prefacEps}. The idea is to smooth out the strong dependence on the fragmentation speed---which is similar to the idea presented in \cite{windmark2012}, where particles have a velocity distribution.

The evolution of the gas is given by \citep{Pringle1981}
\begin{align}
\label{eq:surfgasevol}
 \pd{\surfgas}{t}=\frac{3}{r}\pd{ }{r}\left[ r^{1/2}\pd{ }{r}\left( \nu\surfgas r^{1/2} \right) \right]+\dot{\Sigma}_\mathrm{w},
\end{align}
where $\dot{\Sigma}_\mathrm{w}$ is a loss term due to photoevaporative winds. 
The photoevaporation model is based on \cite{picogna2019} and described in Appendix~\ref{sec:photoevapModel}. 
For the viscosity $\nu$ we choose the turbulent viscosity according to \cite{Shakura1973} which is the same expression as Eq.~\eqref{eq:DgasDiff}.

\subsection{Temperature Model}\label{sec:temp_model}
In order to calculate the midplane gas temperature, one needs the contribution from radiation (internal and external) as well as from accretion heating. 
From pure radiation heating \citep[e.g.,][his section 2.4.2]{armitage2010} one obtains
\begin{align}
    \label{eq:Trad}
    T_\mathrm{rad}=\left(\frac{\Lstar}{4\pi\sigmaSB r^2}\theta+\Tbg^4\right)^{1/4},
\end{align}
where $\theta\approx\tan{\theta}\approx\hscale/r\approx0.04$ \citep[e.g.][]{chiang1997,PfeilKlahr2019}. We set the background temperature due to external sources to $\Tbg=10\,\mathrm{K}$. 

\cite{gough1981} gives a luminosity evolution of the sun, 
\begin{align}
    \Lstar(t)=\frac{\Lsun}{1+2/5\cdot(1-t/t_\odot)}.
\end{align}
As the age of the sun is roughly $t_\odot\approx4.6\cdot10^9\,\mathrm{yr}$ and our simulations run for a few $10^6\,\mathrm{yr}$, we can make the approximation $$\Lstar\approx5L_\odot/7\approx2.73\cdot10^{33}\,\mathrm{erg\,s^{-1}}.$$ 

For pure accretion heating (i.e. ignoring radiation heating for the moment), and without taking optical depth effects into account and assuming that $T_\mathrm{acc}\propto\cs^2$, the local midplane temperature can be calculated as  \citep{nakamoto1994,Pringle1981}
\begin{align}
    T_\mathrm{acc}=\left(\aturb\surfgas\Okepl\frac{9\kB}{8\mgas\sigmaSB}\right)^{1/3}.
\end{align}
Following \cite{ostriker1963} and \citet[his Eq.~(3.37)]{armitage2010}, we approximate the midplane temperature due to accretion and radiation heating as
\begin{align}
    T=\left[\left(\frac{3}{4}\optdepthR+1\right) T_\mathrm{acc}^4+T_\mathrm{rad}^4\right]^{1/4}.
\end{align}
The Rosseland optical depth $\optdepth$ is approximated by
\begin{align}
    \optdepth_\mathrm{R}\approx\kappa_\mathrm{R}\frac{1}{2}\dustsurf,
\end{align}
where $\kappa_\mathrm{R}$ is the size and wavelength averaged Rosseland opacity \citep{birnstiel2018} that is calculated in every time step based on the local size distribution, and $\dustsurf$ is the column density of all particles except planetesimals. 
We use this accretion heating model only for some further test cases. For the majority of presented simulations in the main text, we stick to radiation heating only, see Eq.~\eqref{eq:Trad}.


\begin{figure*}[bt]
\captionsetup[subfigure]{labelformat=empty}
\begin{subfigure}{.499\linewidth}
\centering
\includegraphics[width=\linewidth]{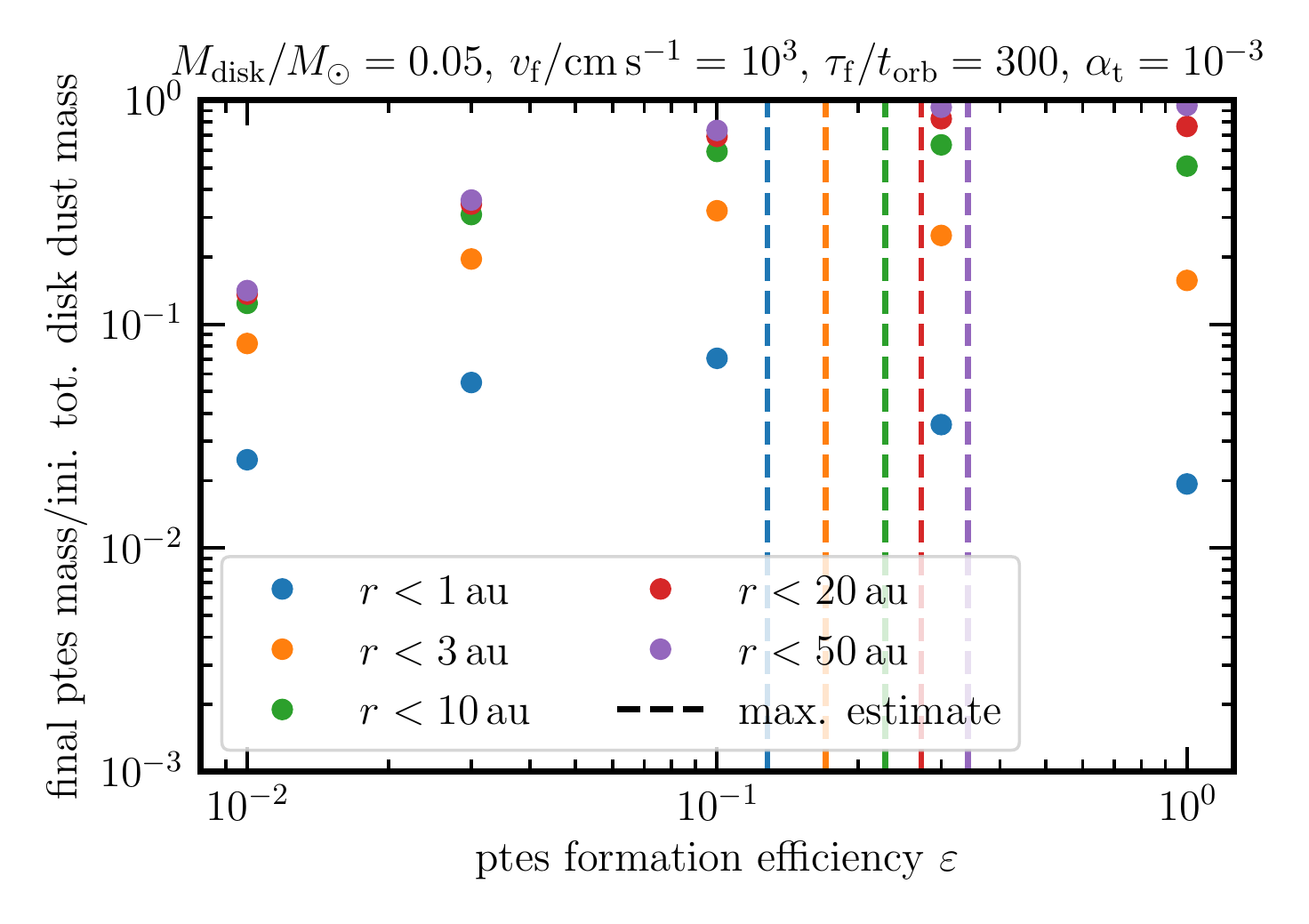}
\end{subfigure}%
~
\centering
\begin{subfigure}{.499\linewidth}
\centering
\includegraphics[width=\linewidth]{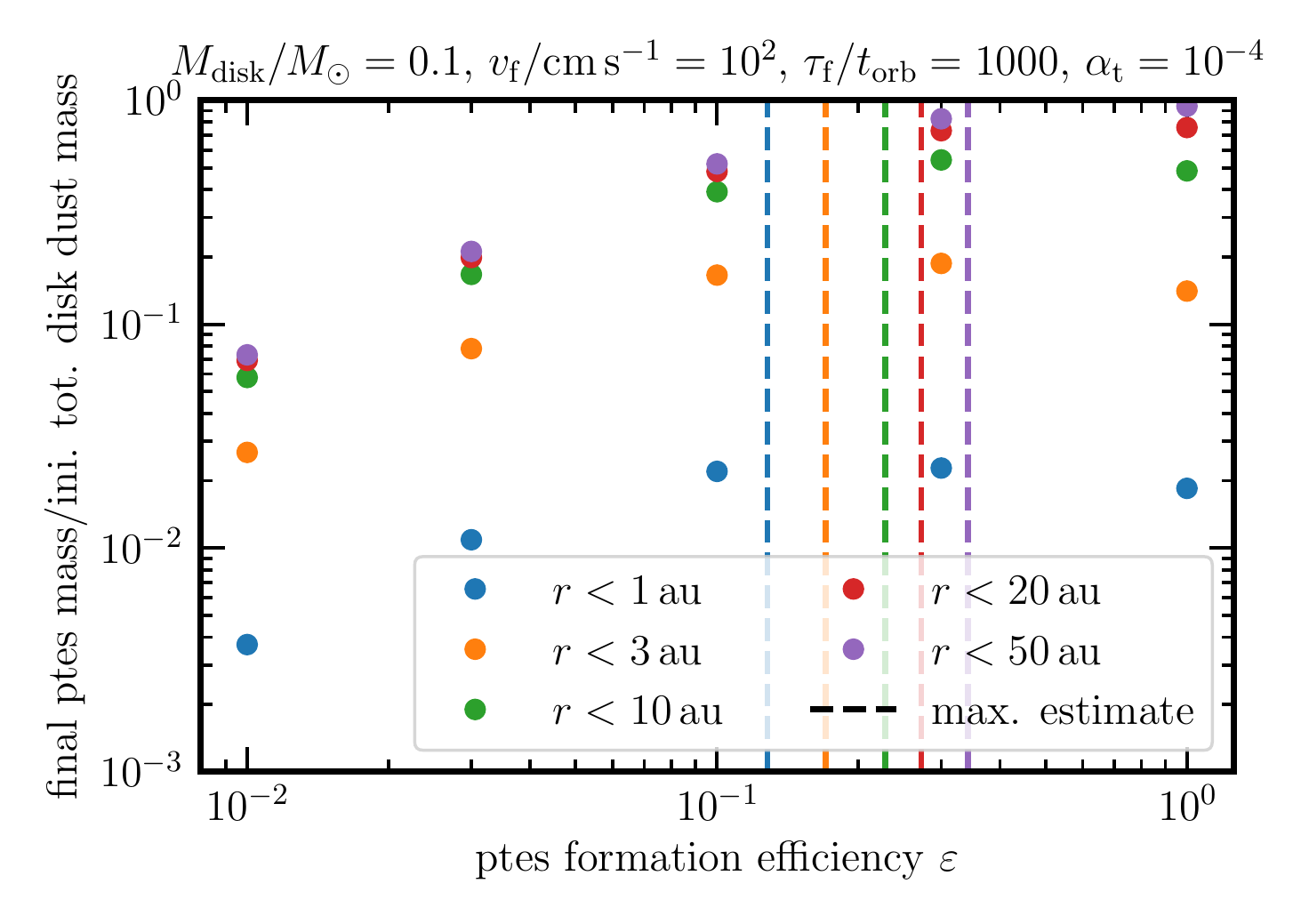}
\end{subfigure}%
\caption{
Final mass in planetesimals within a given disk radius shown in the legend, normalized to the initial total disk dust mass, as a function of planetesimal formation efficiency. Defining $\peff=5\hscale/\ell$, where $\ell$ is the conversion length over which pebbles are transformed into planetesimals. The vertical dashed lines show the predicted maximum at the outer edge of the respective zone. I.e., the blue dashed line shows the predicted maximum at $1\,\AU$ and the purple line at $50\,\AU$. This plot shows $\peff$ variations from the first fiducial run (gray values in Tab.~\ref{tab:paramOverview}) in the left panel and from the second fiducial run in the right panel (bold values in Tab.~\ref{tab:paramOverview}). 
}
\label{fig:BothSamples_epsDependenceMass}
\end{figure*}

\subsection{Analyzed Parameters}
For the total disk masses we used values between the MMSN ($0.013\,\msun$) and roughly the critical value at which disk fragmentation can occur, $\sim0.1\,\msun$ \citep{toomre1964,goldreich1965}. However, for collapse due to the disks own gravity,  the cooling time is also an important criterion \citep{baehr2017}. 

The disk size, which is roughy given by the characteristic radius $\rchar$ of our initial condition, spans from $10\,\AU$ to $100\,\AU$, based on observations \citep{andrews2010}.

For the viscosity power law index $\vispow$ we also allowed extreme cases, i.e. $0.5\leq\vispow\leq1.5$, and made the turbulence parameter disk radius dependent for the cases $\vispow\neq1$:
\begin{align}
    \aturb = \alpha_0 \left( \frac{r}{\rchar} \right)^g,
\end{align}
where $g=\vispow+q-3/2$ and $T\propto r^{-q}$.

For the fragmentation speed, recent work by \cite{musiolik2019} indicated that the value should be around $1\,\mathrm{m\,s^{-1}}$. We still analyze values up to the former default of $10\,\mathrm{m\,s^{-1}}$.

Values for the solar metallicity span from $Z=1.34\%$ \citep{Asplund2009} to $Z=2\%$ \citep[for a review see][]{Vagnozzi2019}. We will use $Z=0.0134$ as our fiducial initial dust-to-gas ratio.

For the trap formation time $\tform$, we took typical  timescales for the significant evolution of disk instabilities --- such as the convective overstability, vertical convective instability, subcritical baroclinic instability, or vertical shear instability \citep{PfeilKlahr2019}, as well as Hall MHD \citep{BaiStone2014,bethune2016}. The values can span from a few hundred to thousands of orbits, depending on how fast the instability evolves and how fast it can then create pressure bumps. Therefore, we should also consider the viscous timescale 
\begin{align}
    \tau_\mathrm{visc}\sim\frac{1}{\aturb\Okepl}=\frac{\torb}{2\pi\aturb}
\end{align}
\citep[e.g.][]{armitage2010} on which structures could form. For example, for $\aturb\sim10^{-4}$ this would give $\sim1600\torb$.

We look at turbulence levels that represent an almost laminar case ($\aturb=10^{-5}$) up to a very turbulent state ($\aturb=10^{-2}$).

For the planetesimal formation efficiency, here defined as $\peff=5\hscale/\ell$, we rely on numerical experiments in order to judge whether values are high or low. We found that $\peff=0.3$ is already high, with almost all the mass that was originially in dust being in planetesimals by the end of the simulations, see Fig.~\ref{fig:BothSamples_epsDependenceMass}. If this ratio is around 0.1, we consider the efficiency to be rather low, which is the case for $\peff\approx0.01$.

X-ray luminosities are found in the range \citep{guedel1997,vidotto2014}
\begin{align}
    10^{28}\lesssim L_\mathrm{X}/(\mathrm{erg/s})\lesssim10^{31}.
\end{align}
Table~\ref{tab:paramOverview} summarizes all the different parameters that we checked for this paper.

\section{Results}
\begin{figure*}[h]
  \centering
  \includegraphics[width=0.75\textwidth]{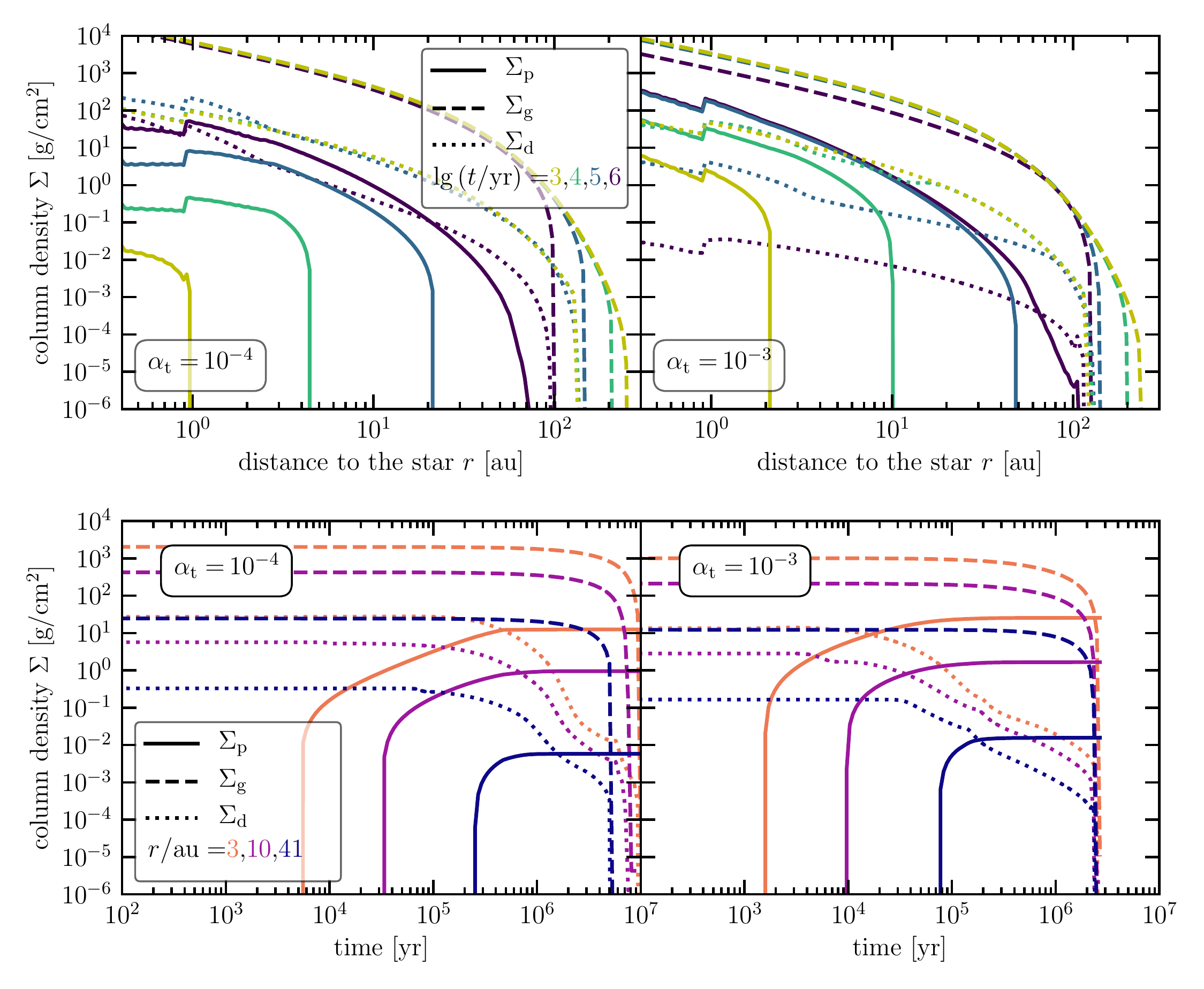}
  \caption
  {\emph{Top panels}: planetesimal (solid), gas (dashed), and total dust (dotted) vertically integrated density profiles at different times. 
  \emph{Bottom panels}: The same quantities, but as a function of time, for three different disk locations.
  Both fiducial simulations are compared, where the first (gray values in Tab.~\ref{tab:paramOverview}) is shown in the right and the second (bold values in Tab.~\ref{tab:paramOverview}) in the left panels.}
  \label{fig:sigma_p_g_d_compareFiducial}
\end{figure*}

\begin{figure*}[th]
\captionsetup[subfigure]{labelformat=empty}
\begin{subfigure}{.45\linewidth}
\centering
\includegraphics[width=\linewidth]{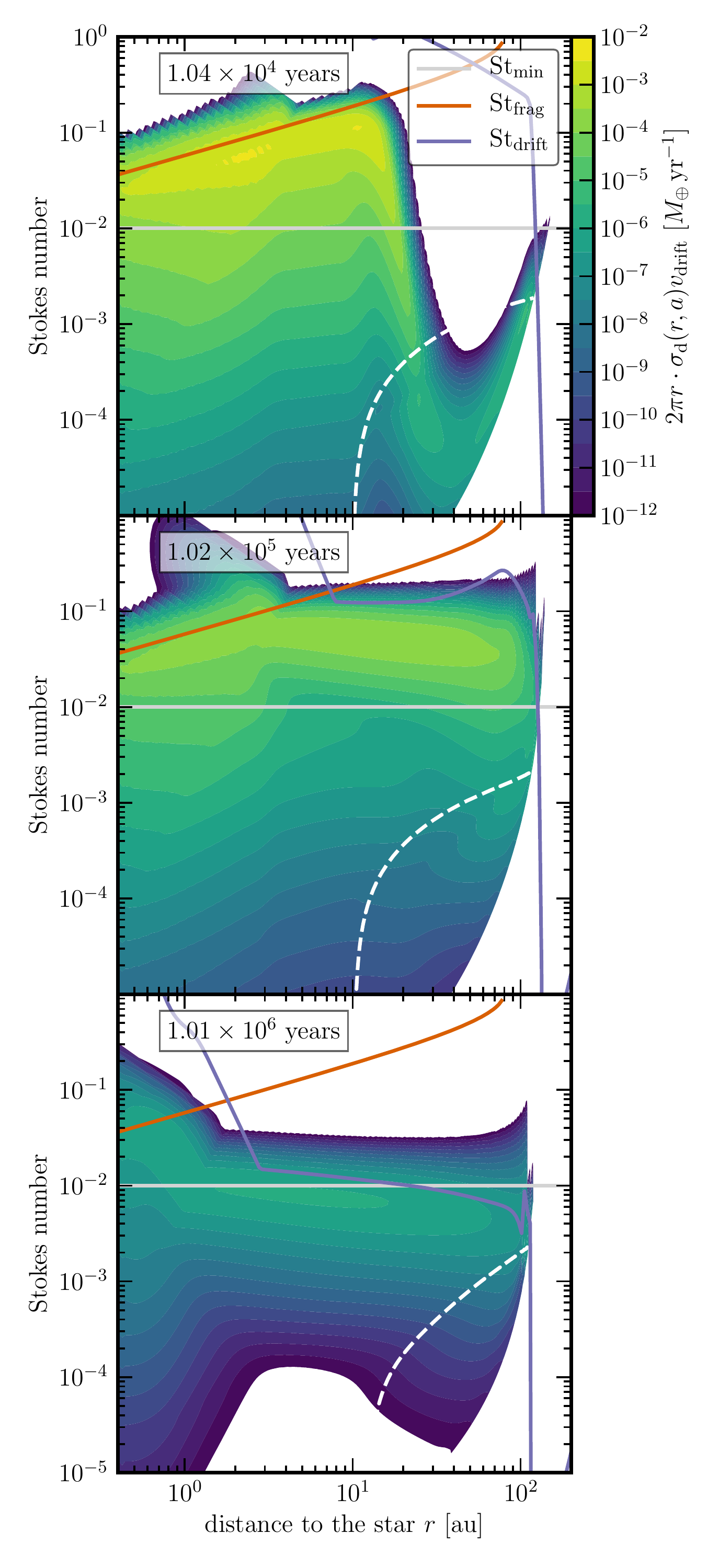}
\end{subfigure}%
~
\centering
\begin{subfigure}{.45\linewidth}
\centering
\includegraphics[width=\linewidth]{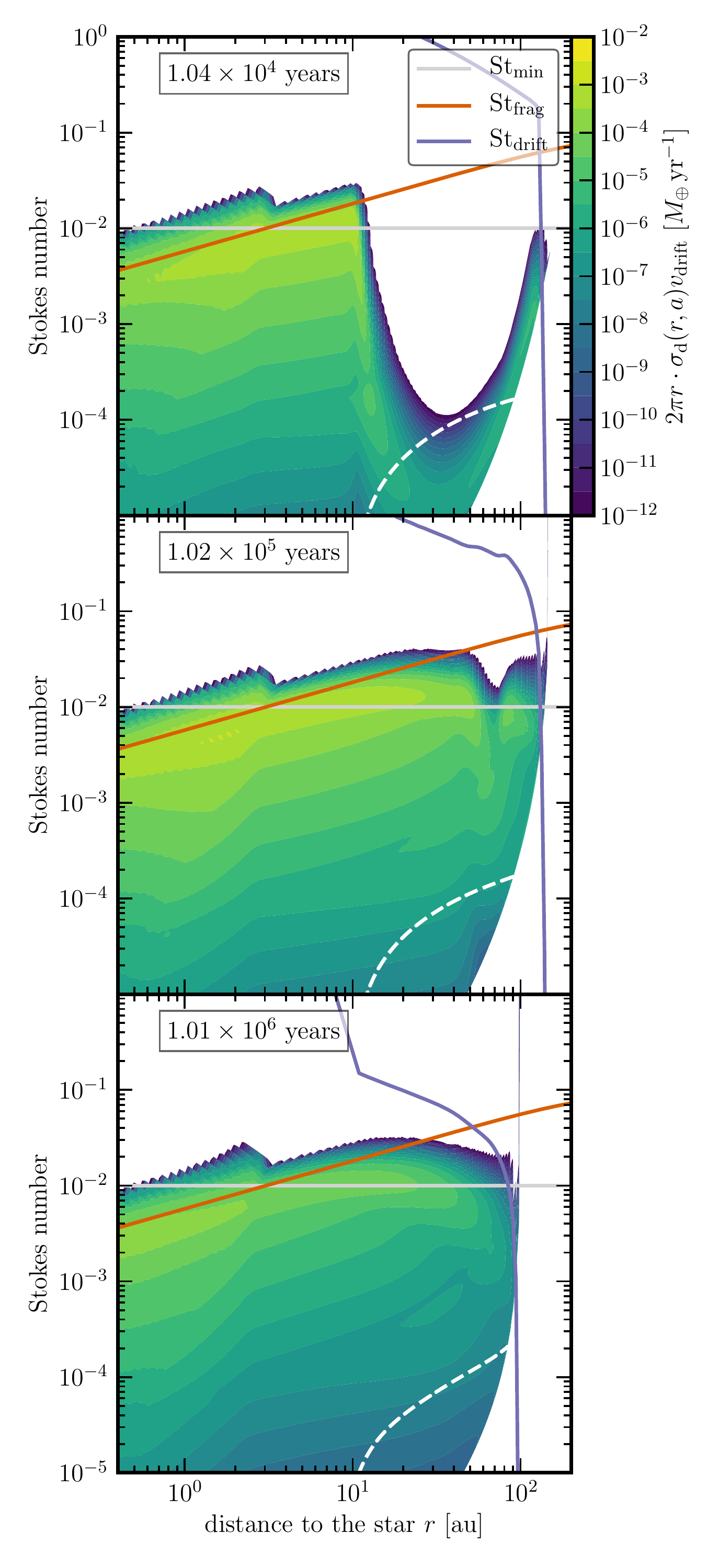}
\end{subfigure}%
\caption{
Local particle flux in Earth masses per year resulting from pure radial drift per size bin (color), as a function of Stokes number and disk radius. Here, we show the first fiducial run in the left panels (gray values in Tab.~\ref{tab:paramOverview}) and the second fiducial run in the right panels (bold values in Tab.~\ref{tab:paramOverview}), both at three different snapshots. The orange (purple) line shows the fragmentation (drift) limit, and the gray line the threshold Stokes number required to participate in planetesimal formation (but see the smoothing function Eq.~\eqref{eq:prefacEps}). Particles in the region below the dashed white line have positive \emph{total} radial velocities, i.e. are moving outward. For the simulation shown in the left panels, outside of $\sim10\,\AU$ the disk is limited by drift over the majority of the time of planetesimal formation. 
For the simulation shown in the right panels, the disk is mostly limited by fragmentation.
}
\label{fig:flux_fiducial1and2}
\end{figure*}
\begin{figure*}[t]
  \centering
  \includegraphics[width=0.8\linewidth]{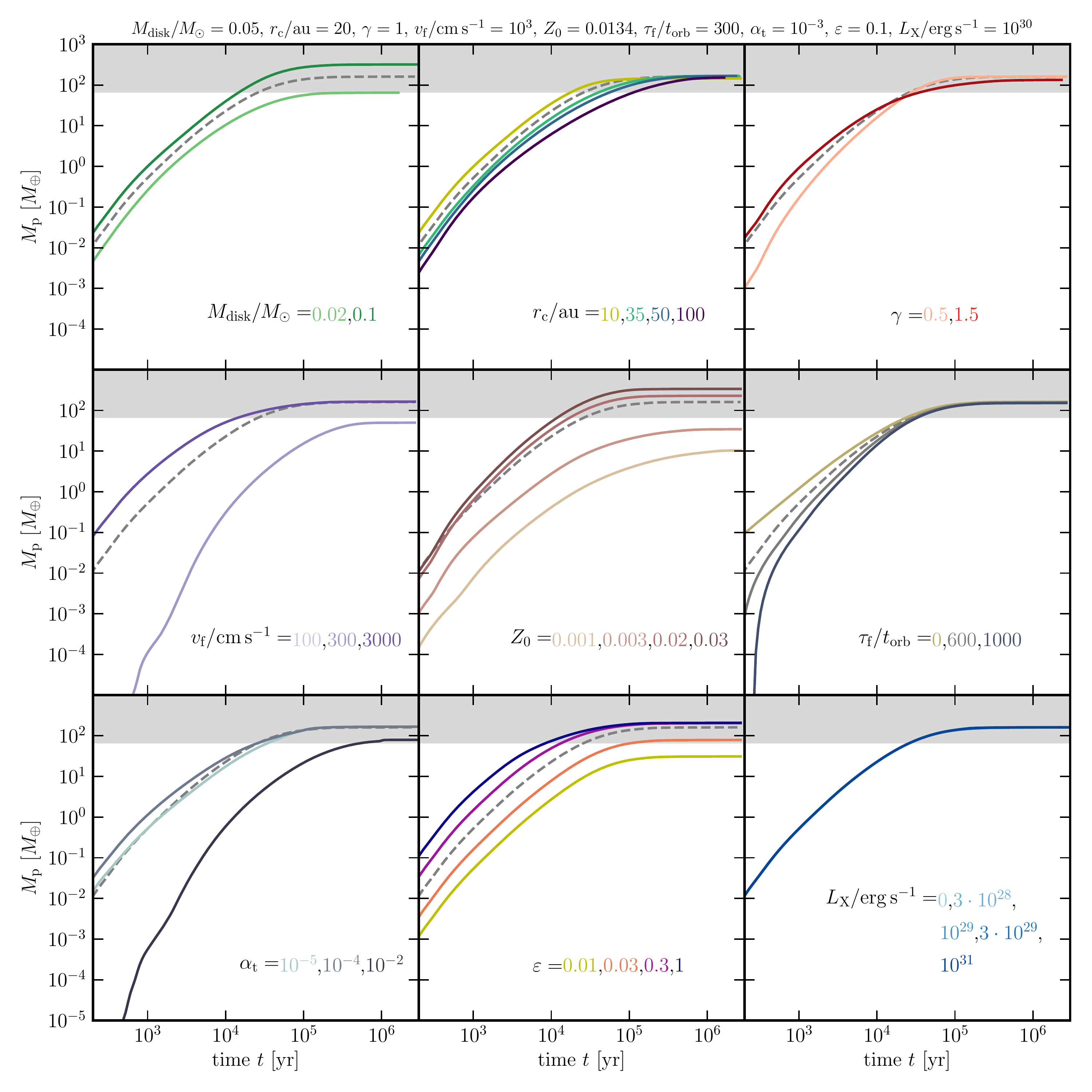}
  \caption
  {Total planetesimal mass as a function of time for the first sample (gray values in Table~\ref{tab:paramOverview}). The fiducial values for this set of simulations are shown in the header of the plot. In each panel, the simulation with those parameters is shown as dashed gray line, and the solid lines with colors show simulation results where only one parameter of the set was changed. The gray area shows values more massive than the solids of the MMSN \citep{weidenschilling1977mmsn,Hayashi1981}.}
  \label{fig:sim_sample_mass}
\end{figure*}

\begin{figure*}[t]
  \centering
  \includegraphics[width=0.8\linewidth]{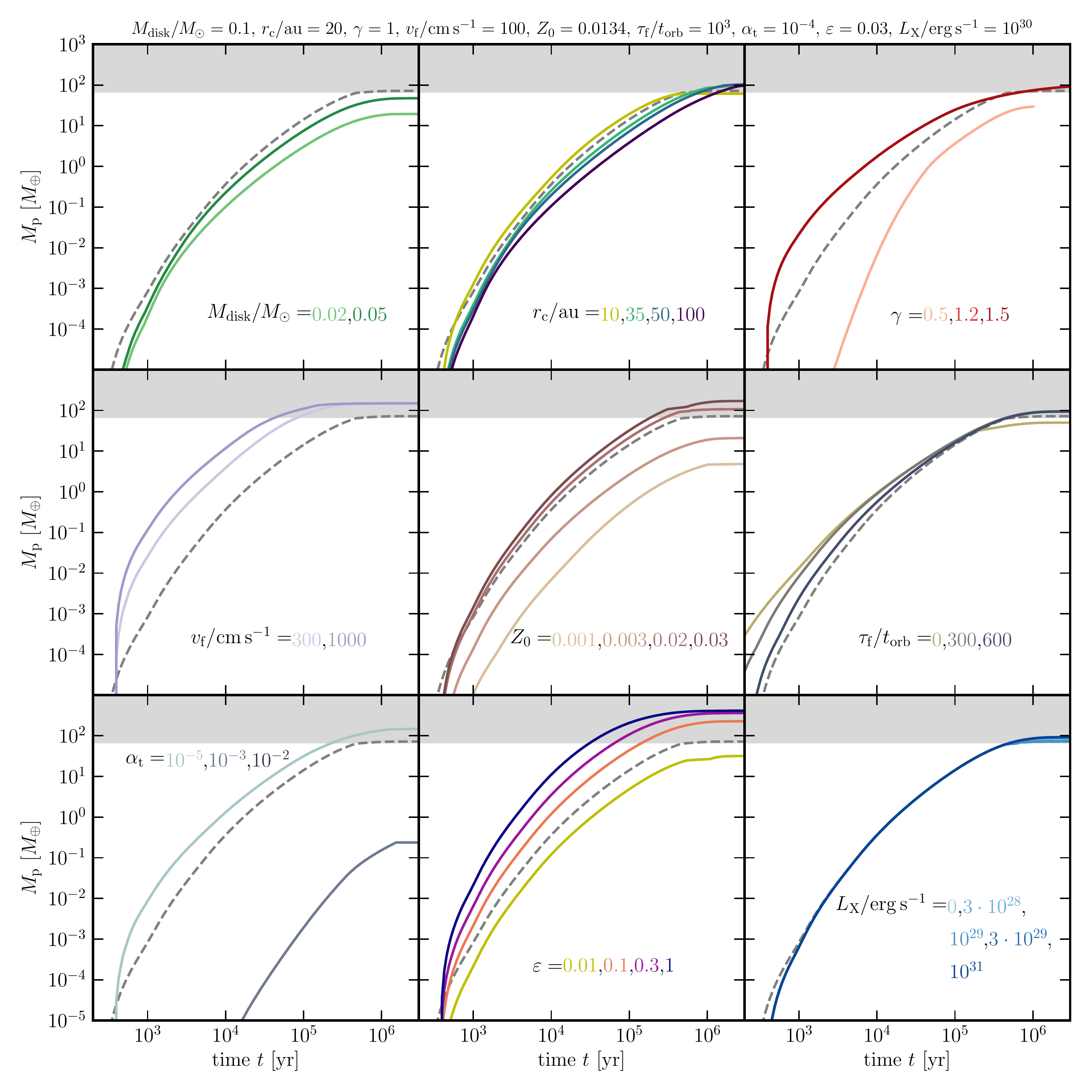}
  \caption
  {Same as Fig.~\ref{fig:sim_sample_mass}, but for the second sample (variations to the bold values in Table~\ref{tab:paramOverview}). The fiducial run produces just enough planetesimal mass in the disk to reach the MMSN mass in solids, even though the disk mass is at the high end already. From this point of view, one needs either a higher initial dust-to-gas ratio, a higher planetesimal formation efficiency, or a larger fragmentation speed. For the bottom left panel, no line is visible for $\aturb=0.01$ as no planetesimals are formed in that case.}
  \label{fig:sim_sample_mass2}
\end{figure*}

\begin{figure*}[t]
  \centering
  \includegraphics[width=0.8\linewidth]{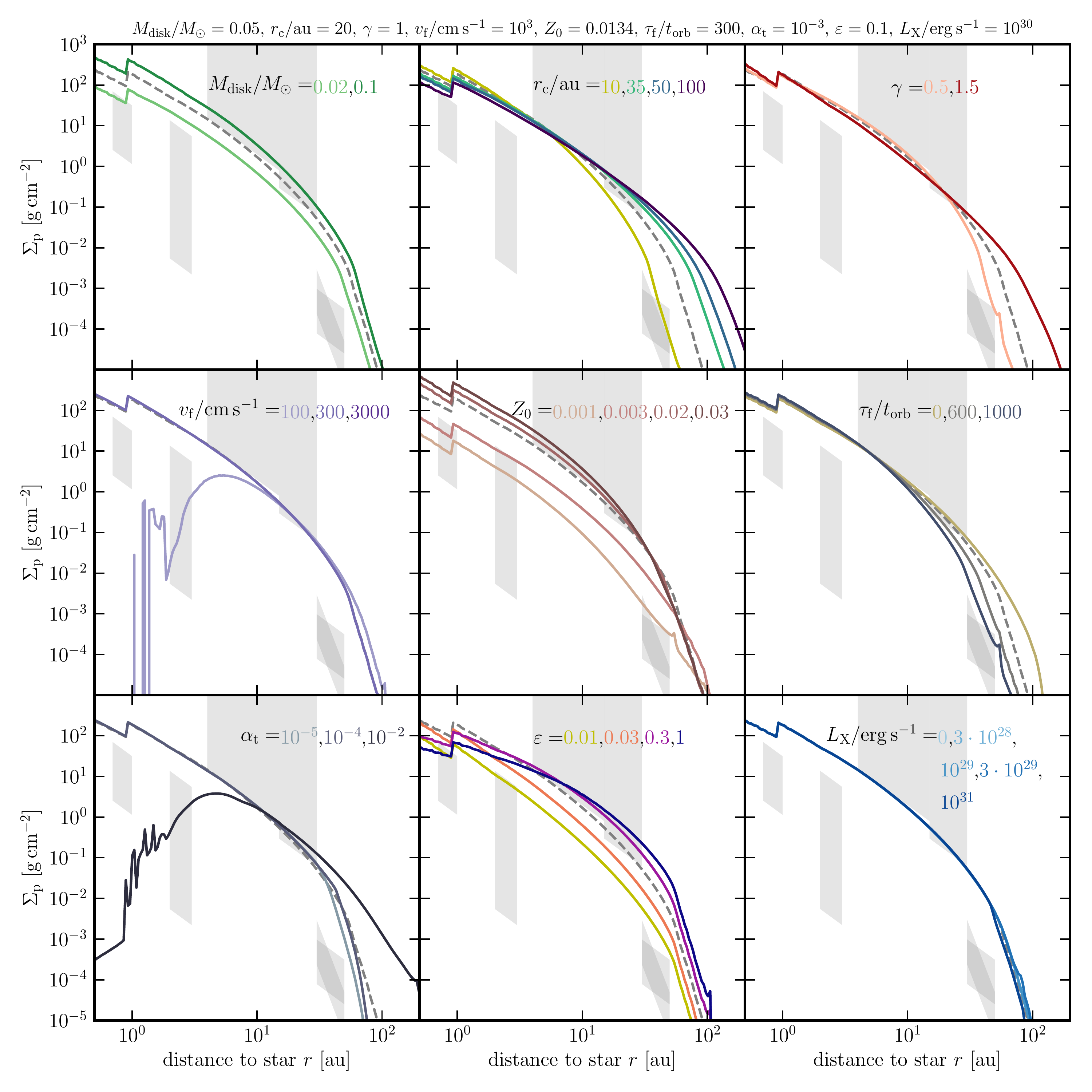}
  \caption
  {Final (at $10\,\mathrm{Myr}$) planetesimal column density as a function of disk radius for the first sample (gray values in Table~\ref{tab:paramOverview}). The fiducial values for this set of simulations are shown in the header of the plot. In each panel, the simulation with those parameters is shown as dashed gray line, and the solid lines with colors show simulation results where only one parameter of the set was changed. The gray areas represent the constraints that we described in sec.~\ref{sec:mass_constr}, where the mass in the given region was translated into a column density, assuming a planetesimal profile $\propto r^{-2.25}$ (see Eq.~(37) of \cite{lenz2019}). For the outermost region, we also overplotted a box $\propto r^{-8}$.}
  \label{fig:sim_sample1}
\end{figure*}

\begin{figure*}[t]
  \centering
  \includegraphics[width=0.8\linewidth]{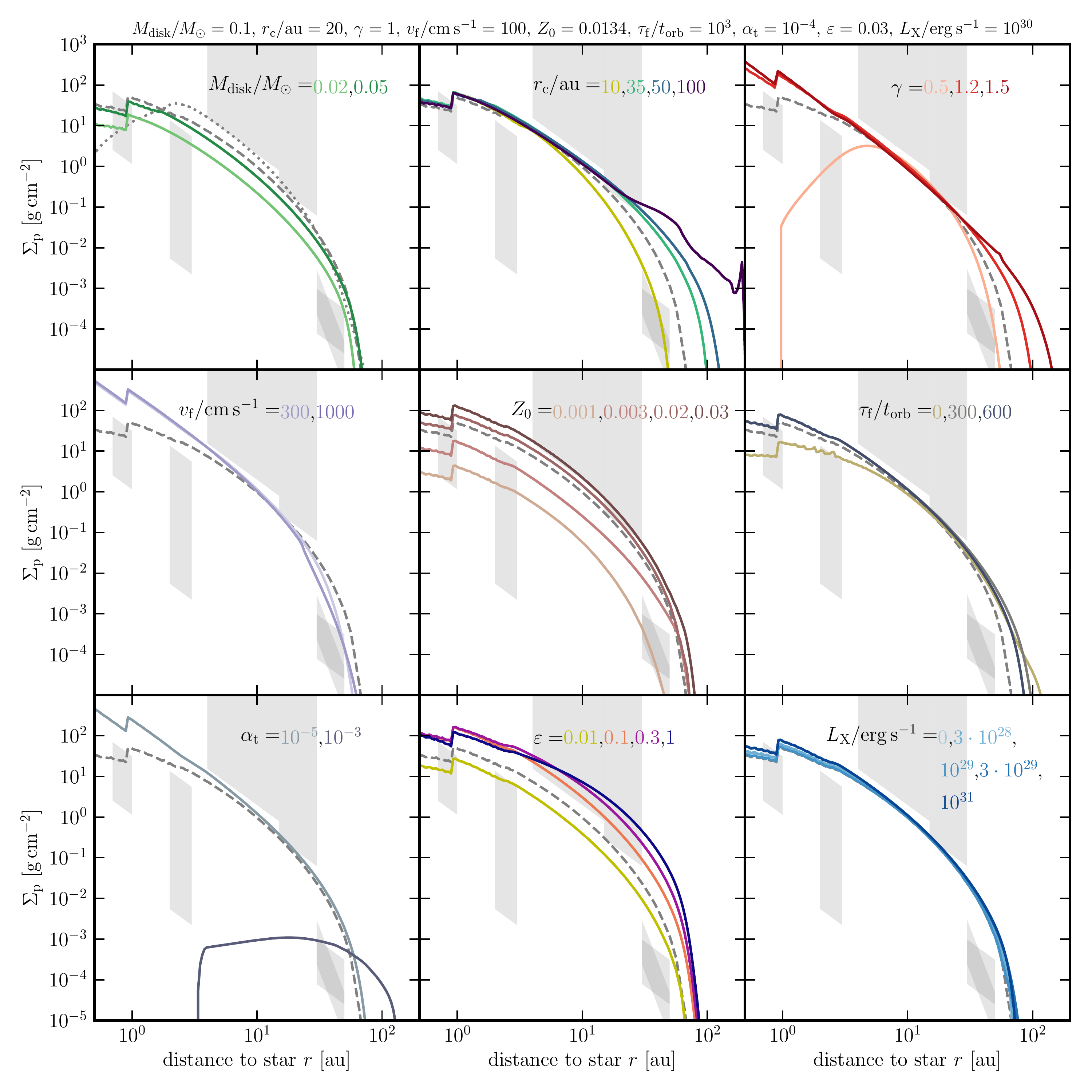}
  \caption
  {Same as Fig.~\ref{fig:sim_sample1}, but for the second sample (variations to the bold values in Table~\ref{tab:paramOverview}). If $\peff$ is high enough, photoevaporation does not change the result significantly. But it seems to have an effect for low $\peff$, see yellow line in the bottom middle panel of Fig.~\ref{fig:sim_sample_mass2}. 
  In the top left panel we also plot the most appealing case as dotted line (last row in Tab.~\ref{tab:paramOverview}).}
  \label{fig:sim_sample2}
\end{figure*}
\subsection{Effect of Planetesimal Formation Efficiency}
Before analyzing the results of all nine different disk parameters, we would first like to concentrate on the planetesimal formation efficiency. Since many processes that we do not fully understand yet are hidden in this parameter, we first have to clarify which values are low or high. 
In the left panel of Fig.~\ref{fig:BothSamples_epsDependenceMass} we show, for the first fiducial parameter set, the final mass in planetesimals within a given disk region over the total initial dust mass as, a function of planetesimal formation efficiency. All simulations were stopped at $10\,\mathrm{Myr}$, or when essentially all the gas was drained. 
One can identify a linear regime for small $\peff$, which makes sense since the formation rate scales linearly with this parameter (Eq.~\eqref{eq:dotsurfp}). For very high efficiency values, the mass in all the regions should reach a plateau, as the planetesimal profile should be very close to the initial condition of the dust. For an extreme case, the conversion length is infinitaly small and pebbles are all instantly transformed into planetesimals. Once a critical large value of $\peff/\trapdist$ is reached, the initial structure is basically reproduced, which leads to a plateau in this plot. Given these extreme cases, we expect there to be a sweet spot, i.e. for a given efficiency the final planetesimal mass reaches a local maximum. 
This  behavior can be described by looking at timescales. 
How fast planetesimals can be built locally from pebbles is determined by the planetesimal formation timescale
\begin{align}
    \tptes  = \frac{\surfpeb}{\dotsurfp}=\frac{\ell}{\vdrift}.
\end{align}
At the same time, pebbles are removed from their location by radial drift, on the drift timescale
\begin{align}
    \tdrift = \frac{r}{\vdrift}.
\end{align}
If $\tptes<\tdrift$, pebbles are transformed into planetesimals faster than particles are removed from their location by radial drift, i.e. the planetesimal profile becomes closer to the initial dust profile. Setting those timescales equal and making use of $\ell=\peff/\trapdist$, one obtains
\begin{align}
    \label{eq:max_mass_peff}
    \peff = \frac{\trapdist}{r}
\end{align}
which means that, for a fixed location in the disk, only the gas temperature matters, since $\trapdist\propto\hscale\propto \sqrt{T}$. Starting from very low $\peff$, the mass increases linearly until the sweet spot where planetesimal formation and drift occur on similar timescales. If $\peff$ is increased even more, pebbles are transformed into planetesimals before they can significantly drift, leading to a profile that is closer to the initial dust profile the higher $\peff$ is.

However, considering the simplicity of the estimate for the maximum, the prediction works surprisingly well for both parameter sets (compare the left and right panels of Figure~\ref{fig:BothSamples_epsDependenceMass}). The difference in the shape of the curves may be related to whether the disk is mostly limited by drift or fragmentation, as well as the mass budget in pebbles. For the fragmentation limited case, material stays within a certain region in the disk longer, as fragmentation events force it to start growing again from tiny, very slowly drifting dust grains, or small dust being swept up by larger grains. As can be seen in Figure~\ref{fig:flux_fiducial1and2}, the high fragmentation speed in the first parameter set allows the disk to be mostly fragmentation limited in the inner disk and drift limited in the outer disk, over the typical time span of planetesimal formation. However, for the second parameter set the fragmentation speed is so low that during the time of planetesimal formation basically the entire disk is fragmentation limited. Also compare to the findings of \cite{Birnstiel2012}, which did not include planetesimal formation. 

The vertical purple lines in Figure~\ref{fig:BothSamples_epsDependenceMass} mark the point beyond which all the initial dust mass has ended up in planetesimals, since there is not much planetesimal mass outside of $50\,\AU$ and since this is a prediction for the maximum in mass at $50\,\AU$. For the radial positions further in, the required planetesimal formation efficiency to reach a maximum in mass is even smaller.

\subsection{Deeper analysis of special cases}
In this section we will focus on the two fiducial runs and the most appealing simulation, where only the latter contains a simple model for accretion heating, that we described in section~\ref{sec:temp_model}. 
Fig.~\ref{fig:sigma_p_g_d_compareFiducial} shows the time evolution of the gas, dust, and planetesimal profiles (top panels) as well as the local values of these at three different disk radii (bottom panels). The kinks in the planetesimal profiles (solid lines) and dust profiles (dotted lines) inidcate the position of the water ice line in that simulation. 
Interior to the water ice line, loss of water ice due to sublimation is assumed.
Fig.~\ref{fig:sigma_p_g_d_compareFiducial} compares both fiducial runs (see gray and bold values in Tab.~\ref{tab:paramOverview}), which have two major differences from each other. 
For the first fiducial run we used a turbulence strength parameter of $\aturb=10^{-3}$ and a fragmentation speed of $\vfrag=10\,\mathrm{m\,s^{-1}}$, whereas the second fiducial run we have set $\aturb=10^{-4}$ and $\vfrag=1\,\mathrm{m\,s^{-1}}$. 
Despite the fact that for the latter case both values are one order or magnitude smaller than in the former, the smaller $\vfrag$ in the second fiducial run leads to much smaller maximum particle sizes, since the fragmentation barrier scales quadratically with $\vfrag$ but only inversely linearly with $\aturb$ \citep{Birnstiel2012}. 
As a result, in the second parameter set almost the entire disk is limited by fragmentation over the major time of planetesimal formation, while the first becomes drift limited much faster (see Figure~\ref{fig:flux_fiducial1and2}).
In Fig.~\ref{fig:sigma_p_g_d_compareFiducial} one can also see that higher $\aturb$ lets the disk spread faster, with material being removed in the very outer regions due to the constant external FUV sink term that we used.

Figure~\ref{fig:flux_fiducial1and2} shows the particle flux for particles of different sizes in the Stokes number space as a function of disk radius at three different snapshots. The fragmentation (red lines) and drift limit (purple lines) set the maximum size of the flux dominating particle species. The horizontal gray line marks the Stokes number beyond which particles are assumed to contribute to particle trapping and planetesimal formation, see also Fig.~\ref{fig:prefacEps}. 
Within $\sim3\,\AU$, some particles have higher Stokes numbers, forming a kink feature because they enter the Stokes drag regime (see Figure~\ref{fig:flux_fiducial1and2} and \ref{fig:flux_aH}). However, it looks much less extreme in the grain size space.  

More details on the special case including accretion heating can be found in Appendix~\ref{sec:specialAccHeat}, where we show the time evolution of the pebble flux and of the planetesimal, dust, and gas profiles.
\subsection{Mass Evolution}
It might be valuable to know if and when the total mass in planetesimals saturates. This saturated mass can be compared with the MMSN solid mass. If this mass is not reached, we consider the parameter set of that simulation to be unable to reproduce the Solar System, as usually several times the MMSN is needed in order to get results that are comparable to the Solar System. 
Figures~\ref{fig:sim_sample_mass} and \ref{fig:sim_sample_mass2} show the time evolution in mass of the first and second sample, respectively. 
In each panel only one parameter was changed compared to the ficucial parameter set shown in the title of both figures. The final value of the total disk mass that is in planetesimals can be compared to the minimum mass for the solar nebula based on \cite{weidenschilling1977mmsn} and \cite{Hayashi1981}. The required mass for initial planetesimals is marked by gray regions. If the final mass is below the gray region for a given parameter set, this set can be excluded for the solar nebula. 
In both figures one can see that initial dust-to-gas ratios of 0.003 and lower are not able to lead to the MMSN mass in planetesimals. A fragmentation speed of $\vfrag=10^2\,\mathrm{cm\,s^{-1}}$ seems to be a critical value, under which the MMSN mass cannot be reached unless the initial dust-to-gas ratio $Z_0>0.0134$, the turbulence strength $\aturb\leq10^{-4}$ (here this parameter is also used for vertical and radial particle diffusion, as well as relative turbulent velocities), or $\peff>0.03$. For a disk starting with $\Mdisk<0.1\,\msun$, $\peff$ or $Z_0$ might need even higher values to reach a final planetesimal disk mass more massive than the MMSN. 

The yellow line in the middle lower panel of Fig.~\ref{fig:sim_sample_mass2} shows a case of low planetesimal formation efficiencies in a mostly fragmentation limited disk. In this case, after about $10^5\,\yr$ photoevaporation allows another phase of planetesimal formation after the planetesimal mass in the disk reached a plateau. The same effect shows up for high initial dust-to-gas ratios, see the case of $\dtog=0.03$ in the centered panel of Fig.~\ref{fig:sim_sample_mass2}. In both cases the reason for the second planetesimal formation phase is the higher mass budget. 
For low $\peff$, particles survive longer in a fragmentation limited disk --- since their average radial drift velocity is much lower, due to disruptive collisions that replenish slowly drifting dust grains --- and less mass is transformed into planetesimals. 
For $\dtog\gtrsim0.03$, the initial particle mass budget is already so high that there remains enough mass for planetesimal formation at later times, when photoevaoration has removed a significant amount of gas mass. However, this effect of a second planetesmial formation phase only occurs if the disk is mostly fragmentation limited, 
which applies for the results shown in Fig.~\ref{fig:sim_sample_mass2} but not for those shown in Fig.~\ref{fig:sim_sample_mass}, in which case the disks are mostly drift limited. 
Additionally, the second planetesimal formation phase induced by photoevaporation is not sufficient to reach the mass of the MMSN for low planetesimal formation efficiencies, i.e. for $\peff\lesssim0.01$.

The first sample, shown in Fig.~\ref{fig:sim_sample_mass}, leads to higher masses of the planetesimal population compared to the second sample,  shown in Fig.~\ref{fig:sim_sample_mass2}. The reason for this is that in the first sample grains can grow to larger sizes due to the higher fragmentation speed. 
However, both samples used the parameterized planetesimal formation model of \cite{lenz2019}, see Section~\ref{sec:ptesFormRate}. 
Again, in this model particle traps are only considered via parameters but the gas profile is smooth, without pressure bumps or gaps, unless caused by photoevaporation. Pressure bumps in the gas profile would lead to a reduction in radial drift speed, allowing particles to remain longer in certain disk regions \citep[e.g.][]{pinilla2012}, even if these traps appear and disappear on a given timescale. This could lead to longer planetesimal formation and greater impact of photoevapotation. However, this might not change the results significantly, leaving the presented conclusions untouched. 

\subsection{Deep Parameter Analysis}
By looking at the final planetesimal profiles for all nine parameters, 
we find a huge variety of possible parameters for the Solar Nebula. It is reassuring that the model works not only for a very finely tuned subset
of parameter choices. 
Though the different parameters can influence each other, it is still possible to draw some conclusions. 
Table~\ref{tab:paramExclude} shows which parameters fail to fulfill the outer Solar System constraints or the MMSN mass. 
In Table~\ref{tab:paramRanges} we present disk parameter ranges that could potentially reproduce the Solar System. Those conclusions are based on Table~\ref{tab:paramExclude}. How much mass the initial disk should contain depends on the fragmentation speed, since the latter determines how much mass is in particles with $\stokes\gtrsim 0.01$. 

Our parameter analysis is based on Figs.~\ref{fig:sim_sample_mass}, \ref{fig:sim_sample_mass2}, \ref{fig:sim_sample1}, and \ref{fig:sim_sample2}. The last two of these figures shows the column density profiles of the final planetesimal population. In each panel only one parameter is varied compared to the fiducial parameter set (dashed lines). In the background, the gray boxes represent the mass constraints discussed in Sec.~\ref{sec:mass_constr}.

The initial characteristic radius $\rchar$ has two major effects. One is the radial position beyond which the dust and gas density drops exponentially. The second is that for smaller (larger) $\rchar$ there is more mass in the inner (outer) disk region. If the disk is too large, there is simply too much mass available around $40\,\AU$ and beyond to form planetesimals. 
Too much mass in the outer disk then leads to violation of the upper CCKBO constraint. Whether this constraint is indeed violated depends also on the initial disk mass, the planetesimal formation efficiency, the initial dust-to-gas ratio, and the viscosity power-law index $\gamma$. However, for a narrow set of these four parameters, finding constraints for $\rchar$ is possible.

The fragmentation speed $\vfrag$ changes the outcome a lot since the fragmentation limit depends quadratically on this parameter. 
For $\vfrag\gtrsim10\,\mathrm{m\,s^{-1}}$ $\vispow\sim0.5$ is actually beneficial due to the stronger density drop in the outer disk, see orange line in the top right panel of Fig.~\ref{fig:sim_sample1}. 
However, for $\vfrag\gtrsim1\,\mathrm{m\,s^{-1}}$ one would need more than $\Mdisk\gtrsim0.1\msun$ to create enough mass to build all planets, specifically in the inner disk part ($\lesssim15\,\AU$), see orange line in the top right panel of Fig.~\ref{fig:sim_sample2}.  

If $\vfrag<1\,\mathrm{m\,s^{-1}}$, too few or no particles with $\stokes\gtrsim0.01$ would be formed, which is the necessary Stokes number to make trapping, and the collapse of pebble clouds into planetesimals work. 
Already, with $\vfrag=1\,\mathrm{m\,s^{-1}}$ it is difficult to meet all the constraints, especially a total mass in planetesimals larger than the minimum mass Solar Nebula \citep{weidenschilling1977mmsn,Hayashi1981} --- see the lowest line in the middle left panel of Fig.~\ref{fig:sim_sample_mass} and the dashed lines in Fig.~\ref{fig:sim_sample_mass2}.

The initial dust-to-gas ratio $\dtog_0$ determines how much mass is initially in particles, and also the dust dynamics, as a low dust-to-gas ratio leads to a drift limited disk that loses particles quickly due to drift. 
Large $\dtog_0$ of up to roughly $0.03$ seem to allow fulfilling the mass constraints on initial planetesimals. However, values of $\lesssim0.003$ lead to too little mass in the final planetesimal population as can be seen in Figs.~\ref{fig:sim_sample_mass}, \ref{fig:sim_sample_mass2}, \ref{fig:sim_sample1}, and \ref{fig:sim_sample2}.

The more time particles have to drift from the region outside of $\sim50\,\AU$ to the inner parts before planetesimal formation, the better the CCKB constraints can be met. 
Alternatively, traps might not occur outside of $50\,\AU$ at all \citep{PfeilKlahr2019}. From the simulations we conclude that the trap formation time must be $\tform>300\,\torb$ or even $\gtrsim1000\,\torb$. 
Other values lead to masses between $\sim30\,\AU$ and $50\,\AU$, which are orders of magnitudes higher than the upper limit.

Constraining values for the turbulence parameter $\aturb$ is also linked to the fragmentation speed, because the fragmentation limit scales inversely linear with $\aturb$ but quadratically with $\vfrag$. For this limit, only the relative velocity matters. However, we assumed that vertical and radial diffusion, as well as the viscosity parameter for the gas, have the same value $\aturb$ that we used for the turbulent velocities. For smaller $\aturb$, particles can settle closer to the midplane. If this value is low enough, growth is not limited by relative turbulent velocties but by relative settling speeds or relative radial drift. When $\aturb$ is high ($\sim10^{-2}$), relative turbulent velocities are too high to allow $\stokes>0.01$ particles. At the same time, the radial viscous gas motion drags dust along to the outer regions of the disk, leading to too much mass in planetesimals outside of $30\,\AU$. Hence, for $\vfrag\sim10\,\mathrm{m\,s^{-1}}$ we suggest $\aturb\sim10^{-5}-10^{-3}$, while the upper end should be up to a few $10^{-4}$ if $\vfrag\sim1\,\mathrm{m\,s^{-1}}$.

Values of $\peff/\trapdist\leq0.002/\hscale$ can be excluded for the solar nebula (if $\peff$ and $\trapdist/\hscale$ are constant), because the total final planetesimal mass in the disk is below the MMSN, and the mass required for the Nice disk can not be reached. Unless the solar nebula was not very small ($\rchar\lesssim10\,\AU$), planetesimal formation should not have been too efficient, i.e., $\peff/\trapdist\lesssim0.06/\hscale$. Otherwise too many planetesimals are formed outside of $30\,\AU$. Thus, the range of possible values is $0.002<\peff\hscale/\trapdist\lesssim0.06$.

Photoevaporation did not influence the final planetesimal profile significantly for small enough disks ($\rchar\lesssim20\,\AU$). However, for large disks ($\rchar\sim100\,\AU$) it can make a difference. But this case is not interesting for finding similar conditions to the solar nebula, as in this case too much mass ends up in planetesimals in the outer disk regions anyway.

In the left panel of Fig.~\ref{fig:BothSamples_epsDependenceMass}, the plateau is reached for smaller values of $\peff/\trapdist$ than in the right panel, which is linked to the smaller pebble mass that is available for planetesimal formation, since the smaller fragmentation velocity leads to smaller maximum Stokes numbers.

\subsection{With Accretion Heating}
Accretion heating leads to a hotter inner disk, which is good since not a single planetesimal was detected inside the orbit of Mercury. I.e., the planetesimal profile has to drop drastically in the inner region before reaching Mercury's current radial position. This can be satisfied due to the higher gas temperatures, as these are forcing the fragmentation limit to be at lower Stokes numbers, and to a larger conversion length ($\ell\propto\hscale\propto\sqrt{T}$). 
In this simulation the ice line is moving radially over time, which is why there is no distinct kink feature in the profile. 
The constraints in the outer disk, that is the Nice disk and the CCKB, are the strongest ones we have. The other constraints may be a bit more flexible. These constraints are roughly met by this simulation.

Since the temperature model presented in this paper was not tested in a comprehensive way, e.g. by comparing with \cite{hubeny1990} or \cite{nakamoto1994}, we only use it to show one special case. In addition, planetesimal-planetesimal collisions would replenish the small dust population \citep{gerbig2019}. This effect is not taken into account in this paper but could change the gas midplane temperature via the mean dust opacity.

We are highlighting one special case with parameters 
$M_\mathrm{disk}/M_\odot=0.1$, $r_\mathrm{c}/\mathrm{au}=20$, 
$\gamma=1$, $v_\mathrm{f}/\mathrm{cm\,s^{-1}}=2\cdot10^2$, $Z_0=0.0134$, 
$\tau_\mathrm{f}/t_\mathrm{orb}=1600$, $\alpha_\mathrm{t}=3\cdot10^{-4}$, 
$\varepsilon=0.05$, and $L_\mathrm{X}/\mathrm{erg\,s^{-1}}=3\cdot10^{29}$. We will refer to this as the most appealing simulation in this paper. The final planetesimal profile is shown in the top left panel of Fig.~\ref{fig:sim_sample2} as a dotted line.
\section{Summary}
We used an extended version of the \cite{lenz2019} model, including Stokes drag for particles, and allowing the gas to evolve viscously while photoevaporation is removing gas over time. The analyzed parameter space was largely increased. While this paper provides a parameter study for pebble flux-regulated planetesimal formation, we focused on meeting Solar System constraints for initial planetesimals. 
Therefore, we used two different default parameter sets and varied one out of nine parameters per simulation. Overall, while some parameters can be excluded, the model seems to be very robust, thus it does not require parameter fine tuning in order to fulfill the constraints. 

The computation times of the presented simulations were between roughly a week and six months, while running on ten cores each. Using more than ten cores for a simulation would not decrease the computation time significantly since the code cannot make use of further parallelization. The runs of the second sample, in particular, were running for months. To shorten the computation time, a simple model must be used such as the two population model presented by \cite{Birnstiel2012}. 
However, this simplified model was only tested for a narrow set of parameters and causes deviations from \texttt{DustPy} simulations for certain parameters. Additionally, the two population model was not yet tested in detail with the inclusion of the planetesimal formation model that was used in this study. 
A simple model reproducing the results shown in this study is likely possible, however, we preferred to use \texttt{DustPy} in order to rely on fundamental physics principles and a sophisticated growth and fragmentation model rather than simplified and untested models.

In sec.~\ref{sec:mass_constr} we suggested mass constraints for initial planetesimals in different regions of the disk: 
    \begin{itemize}
        \item $0.7-1\,\AU$: $0.1-2.77\,\mearth$
        \item $2-3\,\AU$: $0.002-5\,\mearth$
        \item $4-15\,\AU$: $66-\text{unknown}\,\mearth$
        \item $15-30\,\AU$: $10-\text{unknown}\,\mearth$
        \item $30-50\,\AU$: $0.008-0.1\,\mearth$
    \end{itemize}
Within $0.7\,\AU$ and outside of $50\,\AU$ there might have been nothing or a very low mass in planetesimals. These suggested constraints are illustrated in Figures~\ref{fig:mass_constraints_overview} and \ref{fig:density_constraints_overview}.

The fragmentation speed $\vfrag$ that leads to breakup in particle collisions and the turbulence parameter of relative velocities $\aturb$ determine how large particles can grow. If the combination of both leads to a fragmentation limit that is close to Stokes numbers of 0.01, the available mass for planetesimal formation will be affected by these parameters. This is why a constraint in total initial disk mass has to be linked to (mostly) $\vfrag$. To fulfill the constraints we suggested, we need $\Mdisk\gtrsim0.1\msun$ for $\vfrag\sim1\,\mathrm{m\,s^{-1}}$ and $\Mdisk\gtrsim0.02\msun$ for $\vfrag\gtrsim10\,\mathrm{m\,s^{-1}}$. In addition, the solar nebular was not larger than $\rchar\lesssim50\,\AU$ ($\rchar$ is the initial transition radius between a power-law and a dropping exponential profile). The power-law index of that inner region was likely around $\vispow\sim1$, but for large fragmentation speeds $\vispow\sim0.5$ can be beneficial for the outer region due to the density drop (if traps can be formed outside of $50\,\AU$). To allow pebbles with $\stokes\gtrsim10^{-2}$ to form, which is roughly the needed Stokes number for trapping and subsequent planetesimal formation, one needs $\vfrag\gtrsim1\,\mathrm{m\,s^{-1}}$. For the initial dust-to-gas ratio, many values $0.01\lesssim Z_0\lesssim0.03$ could work, but $Z_0\lesssim0.003$ leads to too little mass in planetesimals. Outside of $50\,\AU$ traps needed at least $300\,\torb$, or never formed there. For the turbulence parameter we find a wide range of possible values $\aturb\sim10^{-5}-10^{-3}$ (or only up to a few $10^{-4}$ if $\vfrag\sim1\,\mathrm{m\,s^{-1}}$). Since disk parameters can affect each other, we also find a wide range for the radial pebble to planetesimal conversion length: $0.002<\hscale/\ell\lesssim0.06$. If the disk is sufficiently small ($\rchar\lesssim20\,\AU$), photoevaporation does not change the final planetesimal profile by much. 
    
The parameters of our most appealing case that includes a simple accretion heating model are the following: $M_\mathrm{disk}/M_\odot=0.1$, $r_\mathrm{c}/\mathrm{au}=20$, 
$\gamma=1$, $v_\mathrm{f}/\mathrm{cm\,s^{-1}}=2\cdot10^2$, $Z_0=0.0134$, 
$\tau_\mathrm{f}/t_\mathrm{orb}=1600$, $\alpha_\mathrm{t}=3\cdot10^{-4}$, and
$\varepsilon=0.05$, and $L_\mathrm{X}/\mathrm{erg\,s^{-1}}=3\cdot10^{29}$ (see dotted line in the top left panel of Fig.~\ref{fig:sim_sample2}).
    
We estimated the maximum mass in planetesimals by equating the planetesimal formation and drift timescale. This approach leads to $\ell=r$, see Eq.~\eqref{eq:max_mass_peff}, which seems to fit our simulation results well (see Figure~\ref{fig:BothSamples_epsDependenceMass}). 
If the planetesimal formation timescale is much shorter than the drift timescale, the planetesimal profile reproduces the initial dust profile. Planetesimal formation efficiencies smaller than the value corresponding to this sweet-spot lead to planetesimal profiles steeper than the initial dust profile or even steeper than the minimum mass Solar Nebula profile. This effect was already observed in \cite{lenz2019}, and this study provides an estimate for the transition to more local planetesimal formation, which is linked to slopes closer to the initial dust profile.
    
Within the model, further limitations are that no pebble accretion was included, which could especially affect the planetesimal profile in the inner disk as pebbles get accreted before reaching that zone. In addition, our simulations did not consider planetesimal-planetesimal collisions, which would lead to multiple generations in planetesimals, pebbles, and dust.

\section{Conclusions}
The MMSN is not consistent with viscous disk evolution models and does not provide enough mass in the giant planet forming region to allow strong gas accretion (see Fig.~\ref{fig:density_constraints_overview}). While typically the MMSN distribution is assumed to be present from the beginning, the timing of substantial planetesimal formation could also matter for further embryo formation and evolution. 
We have shown that pebble flux-regulated planetesimal formation produces beneficial planetesimal distributions for a wide range of parameters, both with respect to planetesimal formation and initial conditions of the disk. 
Even though the impact of disk parameters on the evolution of initial planetesimals influence each other, some constraints on these parameters were found in this study. Having only a narrow set of parameters that could reproduce the Solar System would have indicated model fine tuning. 
This stresses the applicability of our parameterization to models of planet formation, e.g. population synthesis models.
\begin{acknowledgements}
C.L. thanks Remo Burn, Thomas Pfeil, Oliver V{\"o}lkel, Giovanni Picogna, Paola Pinilla, Christoph Mordasini, Alessandro Morbidelli, Andreas Schreiber, Bertram Bitsch, Joanna Dr{\k{a}}{\.z}kowska, Vincent Carpenter, Peter Rodenkich, Chris Ormel, Matthew Holman, Matias Garate, and Oliver Schib for helpful discussions. We would like to thank the referee Kleomenis Tsiganis for comments and suggestions on how to improve the readability and quality of the paper. 
This work was funded in parts by the Deutsche Forschungsgemeinschaft (DFG,  German  Research  Foundation) as part of the Schwerpunktprogramm (SPP, Priority Program) SPP 1833 ``Building a Habitable Earth'', priority program SPP 1992: "Exoplanet Diversity" under contract KL 1469/17-1, by the priority program SPP 1385 "The first ten million years of the Solar System" under contract KL 1469/4-(1-3) "Gravoturbulent planetesimal formation in the early Solar System".
Futhermore DFG Research Unit FOR2544 “Blue Planets around Red Stars” under contract KL 1469/15-1. This research was also supported by the Munich Institute for Astro- and Particle Physics (MIAPP) of the DFG cluster of excellence "Origin and Structure of the Universe" and was performed in part at KITP Santa Barbara by the National Science Foundation under Grant No. NSF PHY11-25915. 

\end{acknowledgements}

%
\bibliographystyle{aa} 

%






   
  



\begin{appendix}

\section{Photoevaporation}
\label{sec:photoevapModel}
For the gas loss rate due to photoevaporation we follow \cite{picogna2019} (X-ray and EUV). 
Note that carbon depletion can have significant effects \citep{woelfer2019} which we will not take into account.
For the profile provided by \cite{picogna2019} we used the scaling with star mass from \cite{owen2012}. The equations presented in this section are only for gas, but do not remove particles from the simulation. 
For the sake of brevity, we define
\begin{align}
    x = 0.7\frac{r}{\AU}\frac{\msun}{\mstar}.
\end{align}
The photoevaporation profile is given by
\begin{align}
    \dot{\Sigma}_\mathrm{w}\propto\frac{1}{x^2}\prod_{j=-1}^5 10^{c_j\lg{(x)^{j+1}}}\cdot\sum_{i=0}^5(i+1)\cdot c_i\frac{\ln{(x)}^i}{\ln{(10)}^i}
\end{align}
with parameters
\begin{align}
\begin{aligned}
    c_{-1} &= -\hphantom{1}2.8562 \\
    c_0 &= \hphantom{-1}5.7248 \\
    c_1 &= -11.4721 \\
    c_2 &= \hphantom{-}16.3587 \\
    c_3 &= -12.1214 \\   
    c_4 &= \hphantom{-1}4.3130 \\
    c_5 &= -\hphantom{1}0.5885
\end{aligned}
\end{align}
The expression is normalized such that the total mass loss rate
\begin{align}
    \dot{M}_\mathrm{X}=\int_0^\infty2\pi r\dot{\Sigma}_\mathrm{w}\dif{r}
\end{align}
is given via
\begin{align}
    &\lg{\left(\frac{\dot{M}_\mathrm{X}}{\msun/\yr}\right)} 
    \\
    \nonumber
    &= 
    A_\mathrm{L}\cdot\exp{\left\{\frac{1}{C_\mathrm{L}}\left[\ln{\left(\lg{\left(\frac{L_\mathrm{X}}{\mathrm{erg/s}}\right)}\right)}-B_\mathrm{L}\right]^2\right\}}+D_\mathrm{L},
\end{align}
with parameters
\begin{align}
\begin{aligned}
    A_\mathrm{L} &= -2.7326 \\
    B_\mathrm{L} &= \hphantom{-}3.3307 \\
    C_\mathrm{L} &=  -2.9868\cdot10^{-3}\\
    D_\mathrm{L} &= -7.2580.
\end{aligned}
\end{align}
Outside of $120\,\AU\cdot\mstar/(0.7\msun)$ we set 
\begin{align}
    \dot{\Sigma}_\mathrm{w}\left(r>120\,\AU\frac{\mstar}{0.7\msun}\right)=3\cdot10^{-15}\,\mathrm{g\,cm^{-2}\,s^{-1}}
\end{align}
due to external FUV radiation.

Also for the case of an inner hole, we follow \cite{picogna2019}. 
The hole radius $r_\mathrm{h}$ is implicitly defined via the radially integrated midplane gas number density
\begin{align}
    \int_{0}^{r_\mathrm{h}}\frac{\surfgas}{\sqrt{2\pi}\hscale\mgas}\dif{r}=10^{22}\,\mathrm{cm}^{-2}.
\end{align}
The profile with inner hole becomes
\begin{align}
    \dot{\Sigma}_\mathrm{w,h}\propto
    \frac{a_\mathrm{h}}{2\pi r/\AU}b_\mathrm{h}^{\delta x}{\delta x}^{c_\mathrm{h}-1}
    \left[\delta x\cdot\ln{(b_\mathrm{h})+c_\mathrm{h}}\right],
\end{align}
where $\delta x=(r-r_\mathrm{h})/\AU$ and the parameters are given by
\begin{align}
\begin{aligned}
    a_\mathrm{h} &= 0.11843, \\
    b_\mathrm{h} &= 0.99695, \\
    c_\mathrm{h} &= 0.48835.
\end{aligned}
\end{align}
The gas loss rate is normalized such that 
\begin{align}
    1.12\dot{M}_\mathrm{X}=\int_0^\infty2\pi r\dot{\Sigma}_\mathrm{w,h}\dif{r}.
\end{align}

\section{The case of the most appealing simulation including accretion heating}
\label{sec:specialAccHeat}
\begin{figure}[thb]
  \centering
  \includegraphics[width=0.95\linewidth]{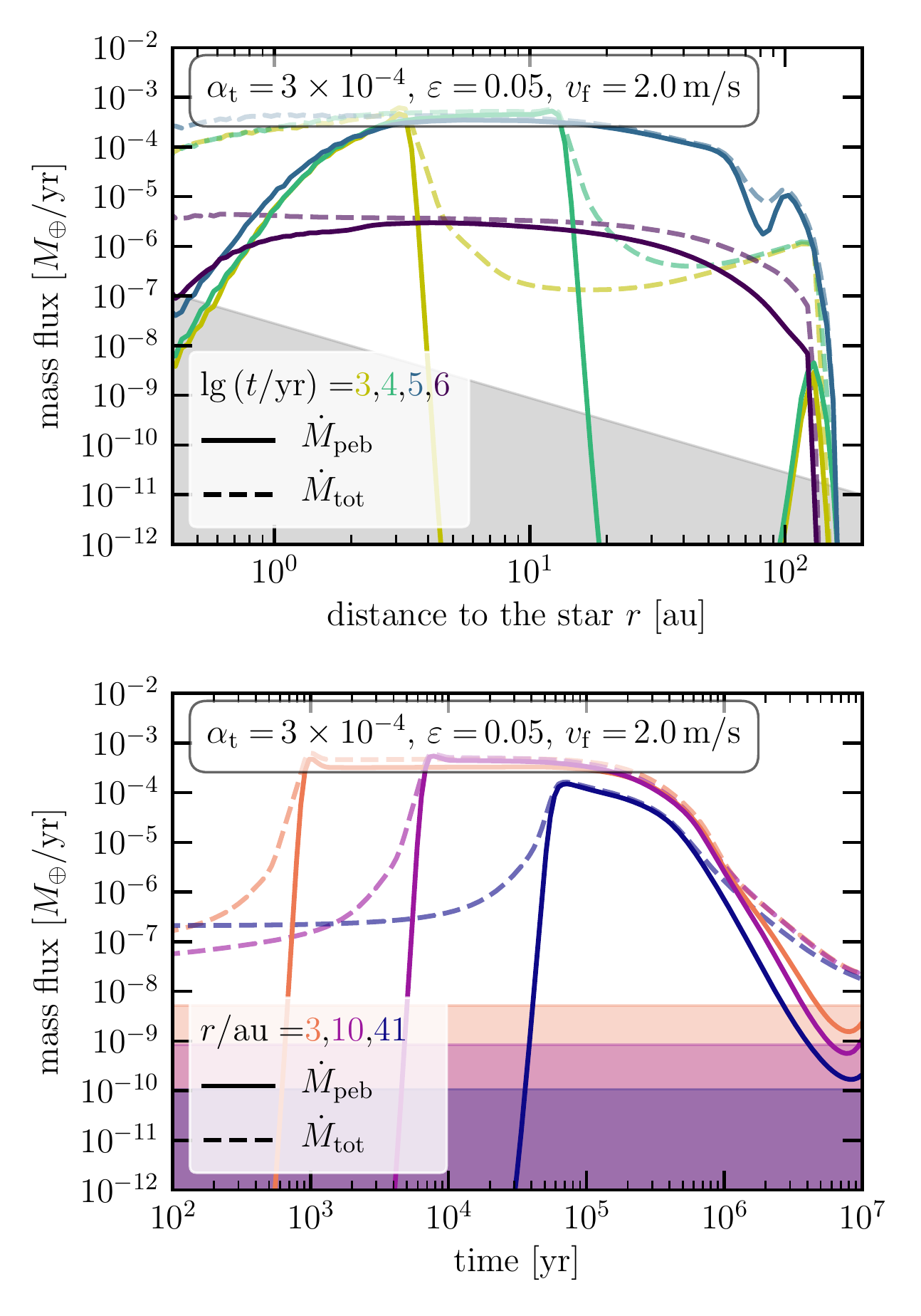}
  \caption
  {Pebble flux in units of Earth masses per year for different times as a function of disk radius (upper panel) and as a function of time for different disk locations (lower panel). Here, we show data from the accretion heating simulation (most appealing case, see last row in Tab.~\ref{tab:paramOverview}). Solid lines show the pebble flux using the smoothing function Eq.\eqref{eq:prefacEps}, dashed lines show the total flux, i.e. taking all solid material into account except planetesimals. For the upper panels, sub-critical fluxes are marked by the gray zone, which are shown in the respective colors in the lower panels.}
  \label{fig:pebbFlux_aH}
\end{figure}

This Appendix concentrates on a special case with accretion heating that is linked to the size distribution of solids. Fig.~\ref{fig:pebbFlux_aH} shows the pebble flux as a function of disk radius at different snapshots and as a function of time at different disk radii. The interpretation of this Figure is similar to the one given in \cite{lenz2019}. 
I.e., once the critical flux for planetesimal formation is reached the flux is orders of magnitude larger than the critical value $\Mcr$ (above the shaded areas in both panels). Photoevaporation (an effect not included in \cite{lenz2019}) leads to a small increase of the pebble flux at late times, see the evolution after $\sim8\cdot10^6\,\yr$ in the lower panel of Fig.~\ref{fig:pebbFlux_aH}. 
However, this increase has only a negligible effect on the final planetesimal population since the pebble flux has dropped by many orders of magnitude compared to its maximum value.

\begin{figure}[t]
  \centering
  \includegraphics[width=0.95\linewidth]{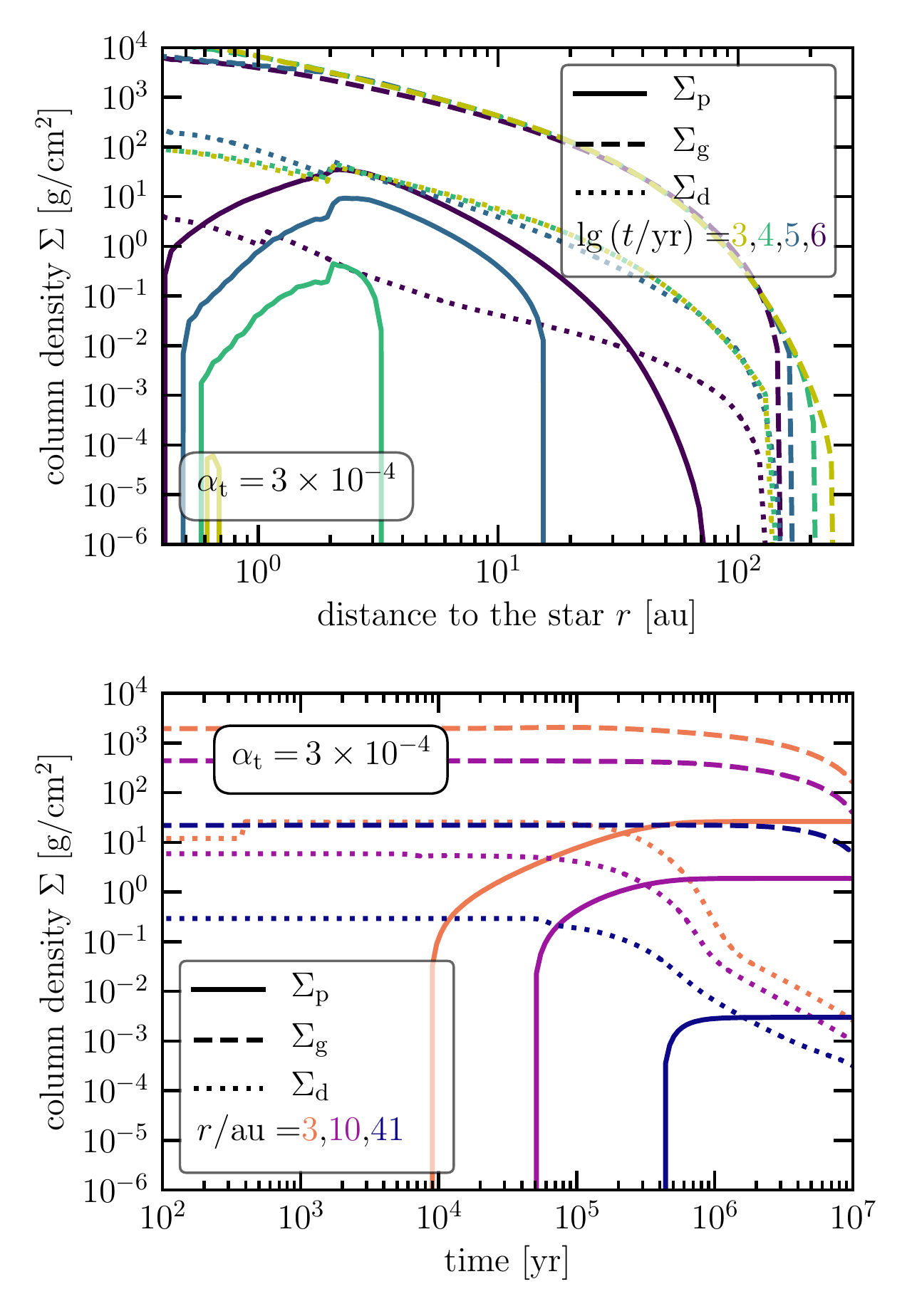}
  \caption
  {Same as Fig.~\ref{fig:sigma_p_g_d_compareFiducial} but for the accretion heating simulation (most appealing case, see last row in Tab.~\ref{tab:paramOverview}).}
  \label{fig:sigma_p_g_d_AccHeating}
\end{figure}

In Fig.~\ref{fig:sigma_p_g_d_AccHeating}, since in this simulation the gas temperature depends on dust evolution, the ice line radially moves over time. Hence, the kink feature in planetesimals that is clearly visible at early times ($\sim10^4\,\yr$) is smeared out at late times ($\sim10^6\,\yr$). At all three locations shown in the bottom panel, planetesimal formation is going on for around $\sim10^6\,\yr$ with significant mass contributions. Note that a higher X-ray luminosity (up to $\sim10^{30}\,\mathrm{erg\,s^{-1}}$) would not change the results by much as the disk would not vanish before $\sim2\,\mathrm{Myr}$. At this time the planetesimal population has saturated already.

\begin{figure}[hb]
  \centering
  \includegraphics[width=0.95\linewidth]{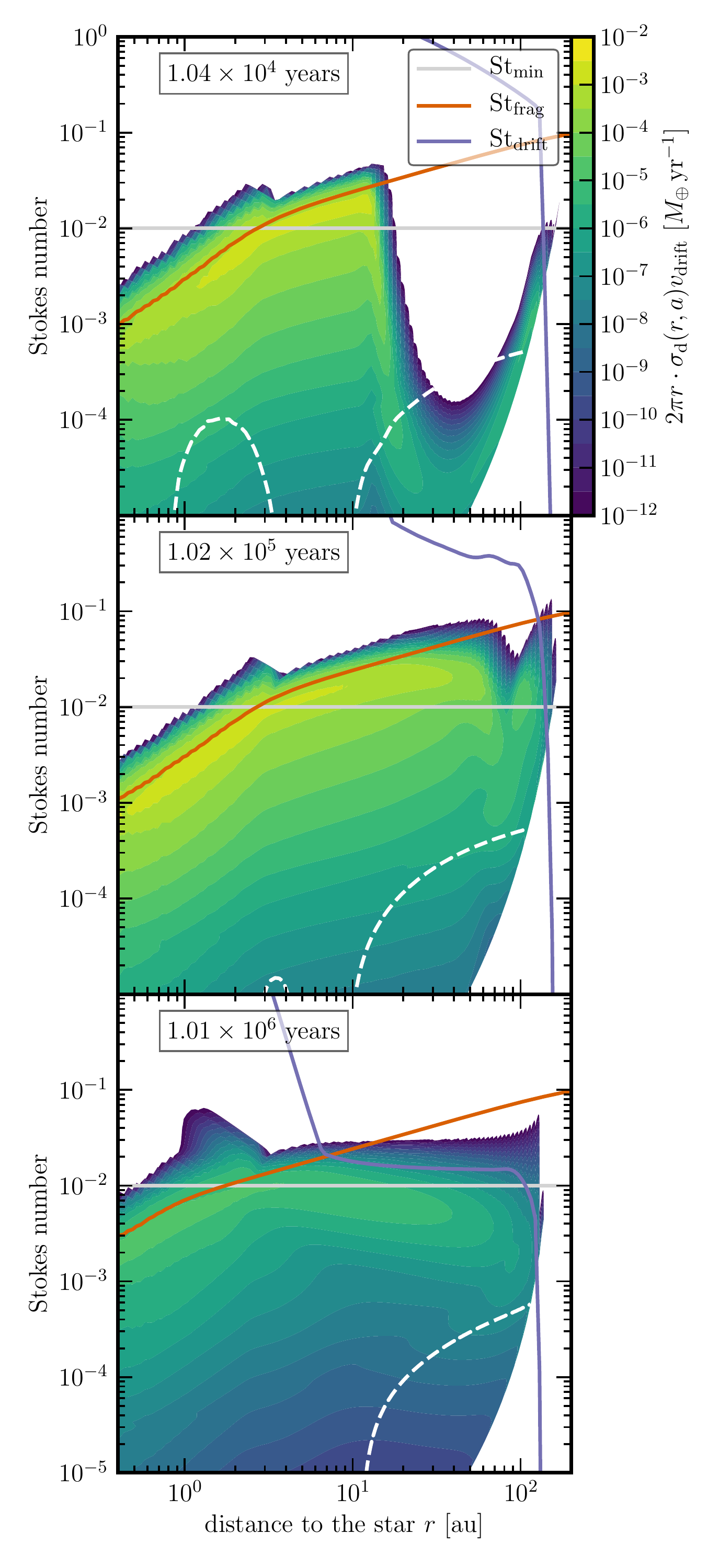}
  \caption
  {As Fig.~\ref{fig:flux_fiducial1and2} but for an example run with accretion heating (most appealing simulation, see last row in Tab.~\ref{tab:paramOverview}).}
  \label{fig:flux_aH}
\end{figure}

For our most appealing case which includes accretion heating (Fig.~\ref{fig:flux_aH}), the higher gas midplane temperatures in the inner disk region are leading to smaller maximum Stokes numbers compared to a situation with pure radiation heating. At late times ($\sim1\,\mathrm{Myr}$), enough dust was converted into planetesimals causing the opacity to drop and thus gas temperatures are much lower than in the initial phase of disk evolution. As a result, higher Stokes numbers can be reached and significantly more planetesimals are formed within $\sim1\,\AU$. This effect is also visible in Figures~\ref{fig:pebbFlux_aH} and \ref{fig:sigma_p_g_d_AccHeating}.
\end{appendix}

\end{document}